\documentstyle[11pt,aasms4]{article}
\def\etal{{\it et al.\ }}
\def\eg{{\it e.g.,}}

\def\putplot#1#2#3#4#5#6#7{\begin{centering} \leavevmode
\vbox to#2{\rule{0pt}{#2}}
\includegraphics{#1}

\end{centering}}

\begin{document}
\title{FAUST Observations in the Fourth Galactic Quadrant\footnote{Dedicated to the memory of Barry Lasker}}

\author{Noah  Brosch{\footnote{Email: noah@wise.tau.ac.il}} }
\affil{The Wise Observatory and 
the School of Physics and Astronomy,
Raymond and Beverly Sackler Faculty of Exact Sciences,
Tel Aviv University, Tel Aviv 69978, Israel}

\author{Alan Brook{\footnote{Email: alan@wise.tau.ac.il}} }
\affil{Department of Geophysics and Planetary Sciences,
Raymond and Beverly Sackler Faculty of Exact Sciences,
Tel Aviv University, Tel Aviv 69978, Israel}

\author{Lutz Wisotzki}
\affil{Hamburger Sternwarte, Gojenbergsweg 112, D-21029 Hamburg, Germany}

\author{Elchanan Almoznino}
\affil{The Wise Observatory and 
the School of Physics and Astronomy,
Raymond and Beverly Sackler Faculty of Exact Sciences,
Tel Aviv University, Tel Aviv 69978, Israel}

\author{Dina Prialnik}
\affil{Department of Geophysics and Planetary Sciences,
Raymond and Beverly Sackler Faculty of Exact Sciences,
Tel Aviv University, Tel Aviv 69978, Israel}

\author{Stuart Bowyer \& Michael Lampton}
\affil{Space Sciences Laboratory and Center for EUV Astrophysics, University of California,
 Berkeley CA 94720, U.S.A.}


\begin{abstract}
We analyze UV observations with FAUST of four sky fields in the general
direction of the fourth Galactic quadrant, in which we detect 777 UV sources. 
This is $\sim$50\% more than detected originally by Bowyer \etal (1995).
We discuss the source detection process and the identification of UV
sources with optical counterparts. For the first time in this project we use ground-based
objective-prism information for two of the fields, to select the best-matching optical 
objects with which to identify the UV sources. Using this, and correlations with
existing catalogs, we present reliable identifications for $\sim$75\% of the sources.
Most of the remaining sources have assigned optical counterparts, but lacking
additional information we offer only plausible identification.
We discuss the types of objects found, and compare the observed population with 
predictions of our UV Galaxy model. 

\end{abstract}


\section{Introduction}

There is a general lack of information about the nature of UV sources fainter than the completeness limit 
of the TD-1 photometric catalog (Thompson \etal 1978). This is, primarily, the
result of a lack of surveys
of the UV sky. Only partial surveys, in limited regions of the sky, have been conducted
since the TD-1 mission. One of the more extended, in terms of general sky coverage, was by the FAUST
experiment. FAUST is a $\sim7^{\circ}.6$ instantaneous field-of-view (FOV) telescope, covering the spectral 
range between 1400 
and 1800\AA\ with an angular resolution of 3.5 arcminutes.  It has a 20 cm entrance
aperture and it utilizes a microchannel plate 
detector with wedge and strip anode, which records the position of each detected photon. 
FAUST operated on board the Space Shuttle (STS) Atlantis in March 1992 as part of the ATLAS--1 mission.
A description of FAUST and its operation aboard ATLAS--1 was given by Lampton \etal (1993).
During this flight 19 pointed exposures were obtained, of which three were short exposures for pointing
checks. The other fixed-pointing images were exposed for 12 to 18 minutes. In addition, two
$\sim$30$^{\circ}$ long scans of the sky were obtained by rolling the STS during the exposure.
Details of the image reconstruction from the time-tagged photon stream and subsequent reductions 
are given in Bowyer \etal (1993).
A catalog of 4660 FAUST sources, originating from the fixed-pointings as well as from the 
sky scans, was produced by Bowyer \etal (1995; FSC). This used a uniform thresholding
algorithm for source detection, and the identifications were obtained from correlations against catalogs 
(mainly from the SIMBAD data base).

The FAUST images are being studied systematically at Tel Aviv in order to identify sources and
perform, as much as possible, ground-based follow-up studies. In this context, we already presented 
results from five images at high Galactic latitude sky regions visible from the Northern hemisphere:
the North Galactic Pole (Brosch \etal 1995), the Virgo cluster region (Brosch \etal 1997), and
in the direction of the Coma cluster (Brosch \etal 1998). This paper presents our analysis of
four FAUST fields in the southern sky. It uses, for the first time, data from an objective-prism
survey (the Hamburg-ESO Survey=HES, Wisotzki \etal 1996; Reimers \etal 1997) to provide additional 
ground-based spectroscopic information about the possible optical counterparts. 

The UV sources in this part of the Milky Way (MW) may help the understanding of the morphology
of the disk and halo of the MW. This is because stellar UV sources represent special stellar 
populations; the majority are rather massive main-sequence objects with spectral types
A \& F, as our previous studies have shown. Their study allows us to test the MW structure in
four spatial windows. In this region there are indications of bends, corrugations, valleys, and 
other morphological peculiarities of the Galactic disk (Alfaro \& Efremov 1996).

The paper is arranged as follows. We first describe the FAUST observations
and the source detection on the UV images. We then describe the process of source identification based on
correlations with published data, mainly from electronic data banks. We describe our use of the HES
objective-prism plates to confirm or modify our proposed identification.
The discussion concentrates on (a) a comparison of the fields which have HES data with those fields lacking 
this information, and (b) the general properties of the UV sources we detected, in the context of the 
Fourth Galactic Quadrant in which they are located. One field is located at the border between the
Third and the Fourth Quadrant; it is analyzed here as though it is one of the 4th Quadrant fields.

\section{The FAUST data and source detection}

The regions observed by FAUST, and in which the source
identification process has been performed, are listed in Table 1. The FAUST images are shown in Figures 1 
to 4, for the  Dorado, Centaurus,  M83, and Telescopium (hereafter Dor, Cen, M83, and Tel) 
regions. All four regions are located in, or very close to, the
fourth Galactic Quadrant, the part of the MW galaxy which (as seen from the solar neighbourhood)
lies between the Galactic center (GC) and l$\simeq$270$^{\circ}$. They are all located at 
intermediate Galactic latitudes, two north of the MW plane and two south of it. The lines of sight sample,
therefore, the disk and halo of the MW at similar ``elevation'' angles from the Galactic
equator and at approximately the same aspect angles with respect to the Galactic bulge. 
Therefore, apart from the general interest in source detection and identification, the
present observations allow, in principle, an inter-comparison of regions which sample 
approximately the same MW environments.

Each FAUST image has location-specific exposure levels within the image. The reason 
is that the Space Shuttle attitude control system permits some platform motion, and the 
information was collected as a time-tagged photon stream. The images were 
reconstructed on the ground by tracking relatively bright stars. This implies that during a pointed 
exposure  a larger
area of the sky was imaged than the instantaneous FOV of FAUST. Table 1 lists, for
each region, the actual sky area covered by each image and the lowest and highest exposure levels, 
generally collected near the edge and in the central part of the image, respectively. The sky
area covered by each image was calculated from the total number of non-zero exposure pixels, and by
assuming that each pixel is a square of 1.1 arcmin sides. This is an approximation,
because electrostatic distortions of the image caused by the detector causes pixels to have slightly
different areas.

An impartial source detector algorithm, described in Brosch \etal (1995), operated on the 
FAUST images  and produced lists of UV source candidates with 
their FAUST UV fluxes and equatorial coordinates. Because of the physical nature of the FAUST frames, 
the exposure times, and therefore the image depths, differ for different regions in each image. Variations 
range from a few hundred seconds in the central regions to a few tens of seconds at the frame periphery. 
This results in regions with reasonably high signal, whereas other regions (with low exposure) have 
low signal levels. The detector used in FAUST has as only sources of noise the Poisson 
statistics of photon arrival and detection, and the sky background. As a result, the detection method
must be signal-to-noise limited rather than flux-limited. The  source detector algorithm used here 
required an acceptance threshold (AT) level of 5.5 times the standard deviation of the local background 
($\sigma_B$) to decide 
that a source was probably real. This was determined by examining the behavior of the
number of source candidates with threshold level; the number of possible sources increased exponentially
with AT for thresholds lower than 5, but only linearly for thresholds higher than this value. 
We infer that, if the AT is too low, large numbers of ``false'' sources are
introduced by the detection method. By adopting a high threshold level one is in danger of 
losing valuable information. We selected the specific 5.5$\times\sigma_B$ level used here at 
the inflexion point in the plot of the number of detected sources (N$_{s}$)
{\it vs.} AT, where the linear regime and the exponential regime cross over. The total number of 
sources detected by this method in each field is also listed in Table 1.

The astrometry of the sources' positions is based on a best-fit to the $\sim$40 brightest UV sources
in each field. These were identified in  predicted UV images, which were created with data 
from the SAO catalog and the Hipparcos Input Catalog (HIC), together with the algorithm for 
predicting the UV magnitude using the visual magnitude and spectral type of stars
(Brosch 1991, and further modifications). The fits were usually good to
$\sim$30" and were checked using the derived coordinates of the remaining $\sim$20 visually identified 
sources. The photon flux from each detected source was measured by integrating all count rates within a 
simulated round aperture centered on the location of the object, and subtracting the 
local background.
A round aperture of diameter 12 pixels, equivalent to 13.2 arcmin, was used for the majority 
of sources except for the very strong sources where a diameter of 24 pixels (26.4 arcmin) was used.
This aperture is much larger than the resolution of FAUST and it allows the collections of
virtually all the UV photons from a source.
The photometric error is composed of the aperture photometry error, that originating from the
background estimation, and a systematic error in flux resulting from the laboratory calibration 
of FAUST and the intrinsic errors of the IUE photometric scale (15.8\% of the observed flux). 
Monochromatic FAUST magnitudes ([UV]=--2.5 log(flux)--21.175) 
are used here to represent the detected brightness, where the flux is in erg/cm$^2$/sec/\AA\,
and was derived from the count rate of FAUST, as explained in Brosch \etal (1995).
The photometric error translates to a minimum error of  0.16 mag in the UV magnitude used here.

The lists of UV sources, separate for each FAUST field, are presented in Table 2. Each source is
identified by a running number and two leading letters, the second that identifies the FAUST field in which
this source was identified. For instance, source FD1 is the first detected UV source in the
Dorado (Dor) field, and FM83 in the 83rd source in the M83 field. We give the J2000 coordinates 
of the sources in columns 2 and 3, the FAUST magnitude and error in column 4, and the
corresponding UV source number from the FAUST source catalog (FSC: Bowyer \etal 1995), for
the cases where the FSC lists a source in column 5.

It is important to emphasize that the source detection algorithm used here is different from
the one used in creating the FSC (Bowyer \etal 1995). Because of this, a comparison of the number of
detected sources listed here and in the FSC is in order. We searched the FSC for all entries within 3' 
of each of the sources detected here. The corresponding FSC number is listed in Table 2 for all entries 
where a match was found. In a few cases more than one match was found. This was usually due 
to the proximity of another source and the ``ownership'' was resolved by visually inspecting the 
UV image. 
For three sources, all in the M83 field, Bowyer \etal (1995) detected two components, while we detected 
only one. The vast majority of the FSC detections have matches in our lists. In the three other fields 
(Dor, Cen, and
Tel) most FSC sources undetected by us are all very faint and remained unidentified in the FSC.
They could be artefacts, but there is one exception: FSC4623 = SAO193381 in the Tel field, 
apparently a bright UV source, whose presence we could not confirm with our detection technique.

A comparison for the M83 field is problematic due to the overlap of one of the FAUST scans with the 
pointed M83 exposure. Nevertheless, the number of FSC entries for the M83 region not detected here is small; 
ten such entries were found within 3$^{\circ}$.4  of the M83 image center. This number does not include the 
three cases where the FSC showed two adjacent entries for only one source detected here. 
In these cases we included both
FSC numbers in the lists, but caution that this may be due to a single  object having slightly different 
FSC astrometric solutions, from the scan and from the fixed pointing. Since almost all of the ten 
other ``undetected here'' 
M83 FSC entries have either nearby detected sources or are very close the edge of the image, 
we conclude that they originate from the FAUST scan, not from the pointed observation. 
We estimate the number of FSC sources we did not detect and which originate 
from the M83 fixed exposure to be at most five. Table 3 summarizes this comparison of 
results and emphasizes the
significant addition of UV sources to those in the FSC by using the present detection algorithm. On
average, FSC sources make up only 2/3 of the UV sources detected by us. Therefore, with our
detection scheme we increased by $\sim$50\% the number of FAUST UV sources in comparison with the FSC.

Figure 5 shows histograms of the UV magnitude distribution in each of the four fields. In general, all
fields show a steady increase in the number of detected sources with increasing apparent magnitude, but
there are some subtle differences. The Cen field, which is at the lowest Galactic latitude, has many more
sources than the other fields. On the other hand, the source distribution with UV magnitude in
the Cen field starts to become incomplete
already at m$_{UV}\simeq$11, approximately 1-2 magnitudes brighter than in the three other fields.

One important feature is noticeable by even a cursory inspection of Table 1, which gives the total number
of detections along with the location of each image and its depth. The M83 image has almost the same number
of detections as has the Tel field, yet the two images differ by a factor of $\sim3\times$ in
exposure depth, with the Tel field being shallower. The discrepancy cannot be due to a higher stellar 
concentration in Tel, because there are no (listed) star clusters in this direction and the
galactic latitudes of the two fields are almost identical. This will be discussed further, below.

\section{Source identification}

\subsection{Using catalog correlations}

The primary identification used on-line data bases, such as the SAO catalog (SAO 1966), the Hipparcos 
Input Catalogue  (Turon \etal 1993), the Tycho catalog (ESA 1997), and the NED
\footnote{The NASA/IPAC Extragalactic Database (NED) is operated by the Jet Propulsion
Laboratory, California Institute of Technology, under contract with the National
Aeronautics and Space Administration.}. The identification criterion was the angular distance: 
the nearer an object is to
the calculated UV source position, the more likely it is to be the actual counterpart. Additional constraints
were imposed on a physical basis. The bluer an optical candidate, the more likely it is to be
the real counterpart. The spectral type of the object, if it was listed in the catalog source,
was an important input to the determination of the accepted counterpart.
Sources for which no physical properties could be found were correlated with Guide Star
Catalog (Lasker, Jenkner, \& Russell 1988) objects, if such could be found in the neighbourhood of a source.

The remaining identifications were extracted mainly from the first edition of
the US Naval Observatory two-colour catalog 
(USNO-A1.0, hereafter USNO: Monet \etal 1996). This catalog is based on scans of the Palomar 
and ESO/SERC Sky Survey blue and red Schmidt telescope plates. The absolute scale of the blue 
and red magnitudes is not accurate, but a 
``local'' comparison of colours between a number of candidates for the same UV source 
is useful for identifying the bluest and 
brightest object,  which is most likely to be the counterpart of the UV source. The second
version of this catalog (USNO-A2.0\footnote{http://vizier.u-strasbg.fr/cgi-bin/VizieR?-file\&-source=USNO2}:
Monet \etal 1998) has improved astrometric positions and photometric
calibration and was used to check a few problematic sources. The photometric improvement 
is based on the B and V magnitudes of the Tycho
catalog for the brighter objects and on CCD observations of standard stars for some sky
regions. However, the final reported photometric accuracy (0.15 mag internal error and up to
0.5 mag due to systematic effects) is not much better that that
of the A1.0 version of the catalog.
The USNO catalog, whose candidates are designated by a ``USNO'' prefix in the lists, 
contains both stellar and non-stellar sources.

On completing the identification process and analyzing the results, a few ($<$10 for all four fields 
combined) fairly strong sources, all very near the edge of the images, remained unidentified. 
As these are relatively strong UV sources they are very probably not artefacts, and a special
effort was exerted to identify them.
An additional search in the Tycho catalog, using a search radius of 5 arcmin, produced good identifications 
for most sources with optical counterparts located within 3-4 arcmin of the UV source. Analysis of other 
identified  sources in the same vicinity showed that a 
systematic positional error appears sometimes at the edge of the images. This is presumably due 
to the electrostatic distortions being the greatest at the edges of the detector, 
and not having enough pre-identified stars in these locations as 
input for the astrometric solution used to compute the celestial coordinates. These additional 
identifications have been included in the lists shown in Tables 2 and 4. 

The entries of Table 4, which reports the identifications, repeat the 
source identifier used here in Table 2 column 1, give  a leading catalog identifier for the 
proposed counterpart (column 2), its V-band magnitude (or B-band, in some cases) in
column 3, a note on its 
spectral type if it is a star or
on its morphological classification if it is a galaxy (column 4), 
its listed B-V colour (or B-R, in some cases) in column 5, 
and the candidate's celestial position in columns 6 and 7. All positions reported here are for J2000.

\subsection{Using data from the Hamburg-ESO Survey}

The HES (Wisotzki \etal 1996) is an objective-prism survey aimed at detecting bright QSOs. It was
performed with the ESO Schmidt telescope using IIIa-J plates and a 4$^{\circ}$ objective prism. The 
plates have a resolution of 15\AA\, at H$\gamma$ (seeing-dependent) and cover the range 3200\AA\, to
5400\AA\,. The celestial area covered by the HES is mainly in the extragalactic sky, roughly above 
$\mid$b$\mid$=30$^{\circ}$. Because of this sky coverage limitation, two of the FAUST regions
analysed here could not be compared with the HES data sets. In particular, we limit the discussion
of the objective-prism spectra to the fields of M83 and of Dorado, as the other two regions have
a galactic latitude too low to be included in the HES.

The HES plates were scanned on the Hamburg PDS with a 30$\times$30$\mu$m$^2$ aperture, and spectra were
automatically extracted using object detection on the Digitized Sky Survey. Each
spectrally observed candidate has, therefore, an astrometric solution good to $\sim$2 arcsec and
a linearized spectrum (using the plate sensitometric relation) covering the range of the HES 
objective-prism plates, which is $\sim$300 pixels long. On average, $\sim$75\% of the spectra do 
not show overlap with other neighbouring spectra and can be used to verify or revise the 
classification and assignation of an optical candidate to a FAUST UV source.

The search for optical counterparts on the HES data set was done by LW in early-November 1998. 
A search in a region 3 arcmin wide around each FAUST source was performed and the candidates 
were extracted and examined
by NB. In a number of instances, the candidate optical object was so bright as to cause 
partial or even full saturation of the objective-prism plate. For several cases, the spectral
information allowing a rough classification could be recovered using additional scan lines
from the two-dimensional scan data. We could thus use the Balmer absorption lines and the 
4000\AA\, Ca II break as classifier criteria in objects as bright as B$\approx$9.5.

An optical candidate was accepted as the probable counterpart of a UV source if it:
\begin{enumerate}
\item was the brightest source, 
\item was the nearest source, 
\item showed strong Balmer absorption lines, 
\item showed a significantly strong blue-UV ``tail'' of the spectrum
\end{enumerate}
Not all these conditions had to be fulfilled simultaneously for a source to be accepted
as a counterpart. We are aware that these may seem to be subjective criteria, but adopt 
them here in absence of a systematic spectral classification of all HES spectra. An example 
of the output produced from the HES plates for this search of optical counterparts, one
page of the many produced for the various candidates and inspected by us,
is shown in Figure 6. Three objects are displayed in this figure; the bottom one is source FM 31
identified as a new sub-dwarf star. The various entries for each candidate include the number of
the ESO-Schmidt plate on which the object appears, various classification parameters, the
image of the field from the Digitized Sky Survey with the candidate object centered, and two 
representations of the objective-prism spectrum, one at full resolution (centered) and
another binned to low resolution (to the right). The strong blue tail, indicative of
a very hot object, is evident for candidate HE1342-2728=FM 31.

The HES plates yield also blue (B$_J$) magnitudes, from the blue part of the spectrum and
a calibration against standard stars (Vanelle 1996). The calibration is good to $\sim$0.2 mag
provided that the spectrum is not overexposed (B$_J\geq$11.5).
In principle, it is also possible to determine the U--B and B--V colour indices of the
candidates from the spectra, using a 
calibration of spectral slope against colour indices; however, here one has to guard even more carefully
against chance partial spectrum overlap.

We emphasize that the assignation of optical counterparts in the Dor and M83 fields was done in a completely
independent way from the assignments based on catalog correlations. In most cases, the HES spectra 
confirmed the original catalog selection of a counterpart.
The counterparts selected from
the HES spectra are listed in Table 4 if they were ``better'' than those originally 
selected using catalog correlations, or if they offered the choice of a single counterpart, 
when the catalog searches
yielded a number of optical sources in the same error ellipse. The HES sources can be recognized 
by the HE prefix identifier
and by the approximate spectral type, derived from the visual inspection of the spectrum
tracing.

Note that in many cases the location of the UV source on the HES contains no object with a
spectrum which could be associated with the FAUST source. In the Dor field 37 UV sources had no
such HES counterpart. In a number of cases, we could not solve the question of the proper
counterpart, even using the HES data. In such cases we reverted to the USNO catalog (version
A2.0) and selected the best counterpart to be the nearest and bluest listed object, and this is the
entry adopted for Table 4.

\section{Results}

In this section we summarize results from the detection and identification process, including the
confirmatory or enhancing results with the HES spectra using the data from Table 4. The summary, which 
splits the source identification into three broad classes (extragalactic, stars, and unknown)
is given in Table 5. One item stands out immediately, that is the low number of ``unknown'' sources
in the Dorado and M83 fields. In these two FAUST fields the percentage of such sources is 19\% and 5\%,
respectively, while it is $\sim$30\% in the other two fields.

The only difference between the first pair of fields and the second is that for Dor and M83 we used
the additional information from the HES survey. This indicates that using the HES resource we could 
classify (in broad spectral types) an additional $\sim$15\% of the FAUST UV source, some of which 
belong to the interesting ``hot, evolved'' stellar variety.
In this context, an important statistic is the number of such sources (sub-dwarfs and white dwarfs)
counted in each field. Whereas the images without HES data have extremely few such sources
(one in Cen and none in Tel), the other two fields contain significantly more sources (nine in Dor and
five in M83). This emphasizes the importance of additional spectral information in
untangling the nature of the (optically faint) hot UV sources.

\subsection{Comparing with predictions}

A comparison of the detected and identified stars
with the predicted number of stars, using the updated model of the UV galaxy of
Brosch (1991). The original model uses the Bahcall \& Soneira (1980)
predictor routines, where the Milky Way is approximated as an exponential disk
and a bulge, each with its own luminosity function. The extinction by dust is treated by
assuming a slab-like distribution of dust along the Galactic equator with an exponential scale
height above the disk. The Bahcall \& Soneira
model yields predictions for optical bands, and it has also been extended to the
near-IR. The UV application of this model uses an optical-to-UV transformation
for stars based on their colours. This was derived from IUE observations and
was extended to late-type stars using Kurucz (1991) model atmospheres. In
addition, the UV predictor model includes the Gould belt, and a thick disk of 
white dwarfs with the scale height determined by Boyle (1989).

The comparison of the observations with the model predictions is shown in 
Figures 7 and 8. We show there the distribution of the differential
star counts (number of stars per UV magnitude) vs. the model predictions in the left
panels, where the actual star
counts are represented by squares with error bars derived from Poisson statistics. 
The right-hand plot for each region
corresponds to the distribution of colours. To produce this, we deleted from the star list
those objects classified as ``hot evolved stars'', such as hot WDs and sd's, based on the
catalog information or on data extracted from the HES spectra, because they are
not predicted well by our model. We also deleted all the extragalactic objects, which
are not included in the predictor model. 

Our model predicts the star numbers and magnitude distributions per square degree, and the 
predicted colour distribution depends on the faint magnitude cutoff. The fainter this cutoff,
the redder is the peak and the median colour of the stellar distribution. Allowing fainter stars in
the calculation of the colour distribution brings in more red objects than blue ones,
driving the total distribution to the red. The actual cutoff in a specific FAUST exposure is
location-dependent. This is because the exposure depth is shallower the closer an object
is to the edge of the FAUST image, the part of the area with the low exposure time. For this 
reason, we adopted the following strategy when
scaling the predictions to the ``effective area'' surveyed:

\begin{enumerate}
\item We selected a magnitude cutoff for the colour prediction at 70\% of the peak in the
magnitude distribution. That is, we set the cutoff to a fainter magnitude than the peak
of the magnitude distribution of the specific field, at the (interpolated) magnitude
level where the stellar magnitude distribution reached 70\% of the peak value.

\item We decided on a scaling constant from the ``per square degree'' prediction of
the model and the real star counts in {\bf both} the magnitude and colour distributions
using the total number of stars detected to be brighter than or equal to m$_{UV}$=11
[N(m$_{det}\leq11$)] and the similar number of stars, brighter than 11 mag, predicted
by the model [N(m$_{pred}\leq11$)]. The ratio of these values was used to scale
both the magnitude distribution prediction and the colour distribution prediction.

\item The actual detected distributions, and those predicted by the model and
scaled as explained above, are shown in Figures 7 and 8.

\end{enumerate}

In deriving the UV-V colour of an object we used the magitude of the counterpart as listed
in Table 4. In many cases, this is the USNO catalog ``blue'' magnitude, which may be
substantially different from the actual V magnitude of the objects. However, we estimate 
that this is a random error, which will only increase the scatter of the results but will
not introduce a systematic bias, as most of the stars that have spectral types tend to be 
A and F, which have small B--V values.

The results, shown in Figures 7 and 8 for the four fields, indicate that the procedure is valid and that the
predicted stellar populations, as well as the general colour distributions, correspond to those 
actually observed. We emphasize that, while the choice of the scaling factor seems arbitrary
by using the total number of stars to 11 mag., the use of the same constant for scaling the 
colour distribution is a constraint, and the subsequent reasonable fit represents an independent 
confirmation of this procedure.

The possible sources of discrepancy between the model predictions and the actual measurements are
presumably stars that are too blue in the UV than the model is able to predict. 
Such stars are, for example, hot evolved objects that are not included in the model.
The discrepancy is emphasized by the 
trend of the plots in the right panels of Figs. 7 and 8 to under-predict the observed colour
distribution at the blue end. This indicates that probably there are more hot evolved stars 
among those observed by
FAUST than revealed by the spectroscopic classification. The assumption is supported by the
size of the under-prediction: the discrepancy is higher for the two fields where we do not have
HES spectroscopic data than in the two fields with such data. This finding indicates that
collections of UV observations have the potential of enlarging the samples of such objects,
with obvious influences on our understanding of the structure and evolution of the
Milky Way.

\subsection{Checking the extremes}

In this section we discuss those stellar sources classified as either hot evolved stars (sd's, 
WDs, etc.), or as red stars (K0 and later), which represent the hot and the cold
extremes of the stellar distribution. It is noticeable that of 24 such objects in the four
FAUST fields treated here, only two are in a image for which we have no HES information.
This demonstrates, again, the value of ancilliary spectral information from which we
extract the special-interest objects.

We identified 15 stars as hot evolved objects, mostly in the Dor and M83 fields. These are
relatively faint UV sources, with a median magnitude of 11.44, and with very blue colours (median
UV-V, or UV-B=--3.65) that imply rather faint optical magnitudes. The optical faintness 
explains the lack of spectral information;
only wide-field surveys using multi-band colours or an objective prism could have picked 
these objects out. Most  are sdO, sdB, or sdA,
but one (FC 89) is classified sdF2 and is indeed very red, with UV-V=3.94. This is a high
proper motion star, as expected of a subdwarf. The source with the most negative colour 
index is FD 46, with UV-V=--5.03.

We searched the lists of objects for those with extremely blue UV-V colours ($\leq$--4)
and found very few of these. Specifically, we detected three such stars in the Dor
field, three in Cen, seven in M83, and two in Tel. Six of these are classified in SIMBAD
or by us as sub-dwarfs, one is an emission-line galaxy, and for the rest we do not have
information beyond the identification with a star-like object and a rough optical
magnitude. It is possible that most are also hot sub-dwarfs or binary systems containing
hot, evolved stars, and spectroscopic follow-ups are necessary to clarify this point.

There are 10 objects classified K0 or later in the four FAUST fields. As expected of
red stars, their UV-V colours are in general large and positive. The colour indices are,
though, much bluer than could be expected  from the effective temperature which corresponds to 
their spectral type. The variations of the colour index within the same allocated spectral
type are considerable; among four K2 stars the spread of UV-V is more than 5.5 mag!
Normally, K stars are expected to have UV-V$\approx$3.2-7.4, but some stars in this class
are more UV-bright than this, presumably because of some form of coronal activity (see \eg \,
Jordan \etal 1987), or because of the presence of a hot secondary.

One object, FM 81, is classified as a M0 star with an expected UV-V$\approx$9.45
colour, yet it shows an almost nil colour index. Checking with SIMBAD, we find this star 
to be RW Hya, a well-known symbiotic system consisting of an M2 giant and a $\sim$80,000K
WD (Kenyon \& Mikolajewska 1995). This demonstrates the advantage of a wide
wavelength base, from the UV to the optical, with which to identify potentially
peculiar objects. Another one, FC 168, is an M3III star with UV-V=+5.4, while
we expect for such an object UV-V=+9.67. Although there are no indications
in SIMBAD of any peculiarity concerning this star, it is likely that this is yet
another binary system, with a hot component responsible for the UV excess.

We remarked in earlier papers that metallic-line A stars (Am) appear fainter in the UV than their
regular counterparts. The present study is particularly useful to check this
behavior because it yielded a relatively large stellar sample. Table 4 lists
six Am stars (earlier than A5) and one late-Bm star. These are plotted  as filled squares in the 
colour-colour diagram shown in Figure 9. The comparison is done with regular A stars
from the same FAUST images. The A0 stars are plotted as circles and the late-A stars 
(A7, A8, and A9) are plotted as filled triangles. We added the reddening vector
assuming the standard Savage \& Mathis (1979) relation as a vector for E(B-V)=0.2
to demonstrate that the discrepancy could not be due to reddening by a standard
Milky Way extinction law.

It is clear that the Am stars are UV-deficient in comparison to most of the A-type 
regular stars, including the latest A stars (A9). This deficiency has been noted before (\eg \, by
van Dijk \etal 1978) and is presumably due to line blanketing of their spectral energy
distributions. Note though that the study of van Dijk \etal, and others dealing with the 
issue of UV deficiency of A stars, concentrated mostly on Ap objects rather than on Am 
stars. In principle, this points a way to use the property  of these
A stars to be less UV-bright than expected from their rough optical spectral classification,
to identify possible candidate Am (or Ap) stars from among those objects in our four fields
for which we found spectral classifications. The requirement is for good UV and optical
photometry of the candidate objects, which should have been previously classified
as A stars.

\section{Galaxies}

Fifty-four objects in Table 4 have been identified as extragalactic objects. In two
instances two optical counterparts could correspond to specific UV sources; the
extragalactic ones are listed in Table 6.

As seen in previous cases, FAUST detected more UV flux than IUE in the few instances where 
such observations are available. In the case of NGC 5253, the IUE spectra yield monochromatic
UV magnitudes of [1482]=11.03 and [1913]=11.39, whereas FAUST measured 9.89$\pm$0.19 mag.
The IUE measurements refer to ``standard'' bands in the SW spectral range,
centered at the listed wavelengths. This discrepancy between FAUST and IUE
is undoubtedly the result of measuring a large object with a small aperture. For NGC 5236=FM 83 
the FAUST magnitude is 7.92$\pm$0.17 whereas IUE measured [1482]=10.48 and [1913]=10.82.
NGC 5135 has [FAUST]=13.84$\pm$1.01 and [1482]=13.94 and [1913]=13.87, presumably the
result of most of the UV emission being produced in the ``composite'' Seyfert 2+starburst 
compact nucleus (Phillips \etal 1983). The amount
of UV flux external to the IUE aperture may be quite significant in some cases.

Among the more interesting extragalactic objects detected by FAUST in these four fields
we note M83 (source FM 70),  a  starburst galaxy in which no less than six
supernovae were discovered in a period of 60 years. A similar case is Tol 34 (source FM 181), which
contains a starburst and a Sy 2 nucleus. Some interacting galaxies appear as UV sources: 
the compact group of NGC 6845 with four peculiar spirals is such an example. Unfortunately, the
angular resolution of FAUST  is not sufficient to show which of the objects is the UV
source. This group would make an interesting target for a future UV imaging observation.

We detected nine UV-emitting galactic nuclei, of which six were identified from the HES spectroscopic data.
One of the more interesting such sources is the optically-violent-variable BL Lac object
EUVE J2009-48.8 (source FT 34). Ours is the first listed UV measurement of
this object, which has been detected in $\gamma$-rays and Xrays, as well as 
in the EUV range. Note that the UV-V colour index we measure is based on the optical 
magnitude listed in the NED. Because of the intrinsic variability of a BL Lac object, 
the real colour index may be different from the ``instantaneous'' one.

Note also the detection of the cD galaxy ESO 444-G046, the largest in the cluster Abell 3558,
as a FAUST UV source.
This object has a neutral colour index, UV-V$\approx$+0.06.
One other object detected by FAUST could belong to the same cluster: this is source
FM 185, a star-like object listed only with its GSC number but which showed emission lines
in HES spectra and appears almost star-like on the blue plates of the HES. The
object was classified by us as an ``emission-line galaxy'' (ELG). The colour index
exhibited by this galaxy is one magnitude bluer than that of the cD galaxy. The ELG
is apparently part of a chain of larger galaxies in the cluster, extending from north-west
to south-east. Lacking a redshift, the assumption that this galaxy is a member of 
the A3558 cluster remains to be proven.

\subsection{FAUST and the Galactic structure}
We remarked above on the puzzling discrepancy in the source count between the M83 and 
the Tel areas. Briefly, the discrepancy is that these two sky regions, at almost identical
$\vert$b$\vert$, have approximately the same number of UV sources but one (M83) has
three times more exposure than the other. The puzzle only deepens when we consider the
normalized source count, that is, the projected source density.

We first define a representative projected source density for a field $i$ as $\sigma_i=\frac{N_{sr}}{A}$, 
where $N_{sr}$ is the number of sources brighter than m$_{UV}$=11 detected in the 
specific field, and $A$  is the angular coverage obtained by FAUST and listed in
Table 1. This magnitude cutoff makes it likely that we are not missing sources
that bright even in the image with the shortest exposure time (280 sec for the central region 
of the Tel image), as indeed the source detection histograms in Fig. 5 show. 

Comparing the four values of $\sigma^*_i=\sigma_i \frac{T_{Tel}}{T_i}$, where 
$T_i$ is the exposure time for the image of area $i$, we find that the Tel regions has a 
$\sigma^*\approx 4.5 \times$
higher than the three other regions, which have almost identical $\sigma^*$ values. 
The Tel region appears to have many more UV sources that it is ``entitled'' to have,
given its projected sky area and depth of exposure.
The discrepancy can {\bf not} be explained by the Galactic model. The modified Bahcall-Soneira 
predictions appear to fit reasonably well both the colour and the magnitude distributions of
UV sources.

One possibility is that the Tel region contains even more of the hot, evolved stellar types 
than do the other three regions. This could, in a way, be understood when considering the
pointing direction: the line-of-sight to the Tel region grazes the bulge of the Galaxy
whereas the three other images have wider separations from the Galactic Center.
If the excess of UV sources in the direction of Telescopium would be due to hot, horizontal-branch
stars in the bulge, these would appear as the sources near the faint completeness limit of the 
FAUST exposure. We estimate this from the colours and magnitudes of zero-age horizontal-branch
models listed in Table 2 of Mould \etal (1996) and from a Galactocentric distance of 8 kpc.

Another possibility could be for the M83 region to have, for some reason, more extinction than the Tel
area. We checked this using the stars identified as A0 in Table 4. There are 15 such stars
in the M83 frame and 11 in the Tel image. The mean B--V values are 0.09$\pm$0.10 and 0.06$\pm$0.07,
respectively. The mean UV--V values, on the other hand, are 0.68$\pm$1.06 and --0.16$\pm$77,
respectively. The difference between the mean B--V values of the two regions could imply an
expected difference in the mean UV--V values of $\sim$0.15, assuming the wavelength
dependence of the extinction as in the ``typical'' galactic law (\eg \, Savage \& Mathis
1979), whereas we measure $\Delta$(UV--V)$\approx$0.84. This could be the case if the
slope of the extinction law would be steeper in the FAUST-observed regions than the
Savage \& Mathis relation, and is supported also by the slope of the UV--V {\it vs.}
B--V relation for all the A0 stars in Table 4. The 43 objects have a regression slope of
6.36$\pm$1.29, whereas the Savage \& Mathis relation indicates a slope of 4.9. A steeper
dependence of the UV extinction may be an indicator for the presence of smaller dust
grains than encountered on average in the Milky Way. We note this as a possibility, but do not
explore it further in this paper.

\section{Conclusions}
  
We presented FAUST UV observations in four sky area, mostly in the fourth
Galactic quadrant. Our detection algorithm identified 777 UV sources in these
regions. We identified the sources with optical counterparts mainly through
correlations with catalogued object, and succeeded in increasing the number 
and quality of 
identifications by including objective-prism spectra. The stellar objects detected and
identified in this program fit reasonably well our model of the UV Galaxy. We
discussed the more interesting stellar and extragalactic sources, and pointed
out that the Tel area may contain horizontal branch stars on the outskirts of the
Galactic bulge.

\section*{Acknowledgements}
This paper made use of the SIMBAD data bank, established and maintained by the Centre 
du Donnes Stellaire in Strasbourg. The paper is dedicated to the memory of
Dr. Barry Lasker, one of the ``founding fathers'' of the Guide Star Catalog and 
of other catalogs derived from scanning of photographic plates.
    UV research at Tel Aviv University is supported by grants from
    the Ministry of Science and Arts through the Israel Space Agency,
    from the Austrian Friends of Tel Aviv
    University, and from a Center of Excellence Award from the Israel
    Science Foundation. NB acknowledges support from a US-Israel Binational
    Award to study UV sources measured by the FAUST experiment. We are grateful to Mr.
Benny Bilenko for updating the model of UV stellar distributions in the Milky Way.
Constuctive remarks by an anonymous referee are gratefully acknowledged.

\newpage

\section*{References}
\begin{description}

\item Alfaro, E. \& Efremov, Yu. N. 1996, Rev.Mex.A.A., 4, 1
 
    \item Bahcall, J.N. \& Soneira, R.M. 1980 Astrophys. J. Suppl  
    44, 73. 

    \item Boyle, B.J. 1989 Mon. Not. R. astr. Soc. 240, 533.

\item Bowyer, S., Sasseen, T.P., Lampton, M. \& Wu, X. 1993,  ApJ,  415, 875

\item Bowyer, S., Sasseen, P.T., Xiauyi, W. \& Lampton, M. 1995, ApJS, 96, 461 (FSC)

\item Brosch, N. 1991, MNRAS, 250, 780

\item Brosch, N., Almoznino, E., Leibowitz, E.M., Netzer, H., Sasseen, T., 
Bowyer, S., Lampton, M. \& Wu, X. 1995,  ApJ,   450, 137

\item Brosch, N., Formiggini, L., Almoznino, E., Sasseen, T., Lampton, M. \& 
Bowyer, S. 1997,  ApJS, 111, 143

\item  Brosch, N.,   Ofek, E.,  Almoznino, E.,   Sasseen, T.,  Lampton, M., \&
Bowyer, S. 1998, MNRAS, 295, 959

\item ESA ``The Hipparcos and Tycho Catalogues'' 1997, ESA SP-1200

\item Hagen, H.-J., Groote, D., Engels, D. \& Reimers, D. 1995, A\&AS, 111, 195

\item Jordan, C., Ayres, T.R., Brown, A., Linsky, J.L. \& Simon, T.
1987, MNRAS, 225, 903

\item Kenyon, S.J. \& Mikolajewska, J. 1995, AJ, 110, 391
 
    \item Kurucz, R.L. 1991 NATO ASI Ser. C., Math. Phys. Sci.,  
    341, 441. 

\item Lampton, M., Sasseen, T.P., Wu, X. \& Bowyer, S. 1993, Geophys. Res. Lett. {\bf 20}, 539

\item Lasker, B., Jenkner, H. \& Russell, J.L. 1988 in ``Mapping the sky: Past heritage and future
directions'' (S. Debarat, ed.), Dordrecht: Kluwer, p. 229.

\item Monet, D. \etal 1996, ``USNO-A1.0 catralog'', Washigton: U.S. Naval Observatory

\item Monet, D. \etal 1998, ``USNO-A2.0 catalog'', Washigton: U.S. Naval Observatory

\item Mould, J.R., Watson, A.M., Gallagher, J.S. III, Ballester, G.E.,
Burrows, C.J., Casertano, S., Clarke, J.T., Crisp, D., Griffiths, R.E., 
Hester, J.J., Hoessel, J.G., Holtzman, J.A., Scowen, P.A., Stapelfeldt, K.R.,
Trauger, J.T. \& Westphal, J.A. 1996, ApJ, 461, 762

\item Paturel, G., Vauglin, I., Garnier, R., Marthinet, M.C., Petit, C., 
Di Nella, H., Bottinelli, L., Gouguenheim, L. \& Durand, N.  1992
{\it LEDA: The Lyon-Meudon Extragalactic Database}, CD-ROM Observatorie de Lyon

\item Phillips, M.M., Charles, P.A. \& Baldwin, J.A. 1983, ApJ, 266, 485

\item Reimers, D. \& Wisotzki, L. 1997, The Messenger, 88, 14

\item SAO 1966, ``Smithsonian Astrophysical Observatory Star Catalog'', Washington: Smithsonian 
Institution, publication No. 4652

\item Savage, B.D. \& Mathis, J.S. 1979, ARAA, 17, 73

\item Thompson, G.I., Nandy, K., Jamar, C., Monfils, A., Houziaux, L., 
Carnochan, D.J. \& Wilson, R. 1978, {\it Catalogue of Stellar Ultraviolet 
Fluxes}, The Science Research Council (TD-1)

\item Turon, C. \etal 1993 Bull. Inf. CDS 43.

\item van Dijk, W., Kerssiers, A., Hammerschlag-Hensberg, G. \& 
Wesselius, P.R. 1978, A\&A, 66, 187

\item Vanelle, C. 1996, MSc thesis, Hamburger Sternwarte

\item Wisotzki, L., Kohler, T., Groote, D. \& Reimers, D. 1996, A\&AS, 115, 227

\end{description}

\newpage

\section*{Figure captions}

\figcaption {The FAUST image of the Dorado field, corrected for vignetting and 
effective exposure times, are shown
as grey scale images where the scaling is from the count rate.}

\figcaption {The FAUST image of the Centaurus field, corrected for vignetting 
and effective exposure times, are shown
as grey scale images where the scaling is from the count rate.}

\figcaption {The FAUST image of the M83 field, corrected for vignetting 
and effective exposure times, are shown
as grey scale images where the scaling is from the count rate.}

\figcaption {The FAUST image of the Telescopium field, corrected for vignetting 
and effective exposure times, are shown
as grey scale images where the scaling is from the count rate.}

\figcaption {Distribution of UV magnitudes (in the FAUST band) for the four FAUST fields. These histograms
show all the detected sources, with no attempt to separate stars from galaxies. We label the
horizontal axis, which represents the UV magnitude, as ``FAUST'' to emphasize that
these are monochromatic magnitudes in the FAUST band, as explained in the text.}

\figcaption {An example of the output produced from the HES plates to serve the search
for optical counterparts. The page shows three objects: HE1342-2736, HE1342-2735, and HE1342-2728.
the latter is the chosen counterpart for the UV source FM 31, and the HES spectrum
identifies it as a hot sub-dwarf.}

\figcaption {Distribution of UV magnitudes and colours for the Dor and Cen fields, compared
with the predictions of our Galaxy model. The objects depicted here are only those not
identified as extragalactic objects or as hot, evolved stars (sd's or WDs).}

\figcaption {Distribution of UV magnitudes and colours for the M83 and Tel fields, compared
with the predictions of our Galaxy model. The objects depicted here are only those not
identified as extragalactic objects or as hot, evolved stars.}

\figcaption {Regular and metallic-line A stars in a colour-colour diagram. The Am stars are represented
by filled squares. The normal A0 stars are plotted as empty circles and the
normal A7-A9 stars are represented by filled triangles. The vector originating from
the (-0.1, -1) corner represents the trajectory of a point produced by reddening with
E(B-V)=0.2 mag, using the Savage \& Mathis (1979) Galactic extinction law.}


\newpage
\begin{deluxetable}{cccccccccc}
\tablecaption{FAUST fields analyzed in this paper }
\small
\tablehead{\colhead{Name} & \colhead{$\alpha$(J2000)} & \colhead{$\delta$(J2000) } & 
\colhead{l} & \colhead{b} & \colhead{Exp.(sec.)}  & \colhead{Coverage (deg$^2$)} &
\colhead{ N$_s$} & \colhead{N(m$\leq$11)} & \colhead{Cutoff}}
\startdata
Dor & 04 15 36 & -56 04 12 & 266.0 & -43.7 & 200-610 & 60.5 & 142 & 58 & 12.7 \nl
Cen & 12 47 24 & -40 46 12 & 302.1 & 22.1 & 300-1100 & 62.0 & 249 & 129 & 12.5 \nl
M83 & 13 34 48 & -28 45 00 & 314.3 & 33.2 & 250-850 & 60.1 & 197 & 81 & 13.0 \nl
Tel & 20 11 38 & -47 06 00 & 352.5 & -32.8 & 100-280 & 52.6 & 189 & 105 & 12.35 \nl
\enddata

\tablecomments{The exposure times listed are for regions close to the border of the image, and
at the image center where the exposure is deepest. N$_s$ is the total number
of sources detected in the specific image, N(m$\leq$11) is the number of stars that are
not known to be hot, evolved objects, and the column labeled ``Cutoff'' gives the magnitude
at which the number of stars per magnitude bin drops to 70\% of the peak.}

\end{deluxetable}

\begin{deluxetable}{ccccc}
\tablecaption{Comparison of present detections and the FSC}
\small
\tablehead{\colhead{Name} & \colhead{N$_{detect}$} & \colhead{FSC} & 
\colhead{FSC-U}  & \colhead{FSC-U'} }
\startdata
Dor & 142 & 87  & 1  & 0  \nl
Cen & 249 & 164 & 4  & 1 \nl
M83 & 197 & 142 & 4a & 2a \nl
Tel & 189 & 120 & 9  & 2 \nl \hline
Total & 777 & 513 & 18 & 5 \nl
\enddata

\tablecomments{FSC-U is the number of FSC sources that are not at the
edge of an image, and
FSC-U' is the number of sources that are not at the edge of
an image and we could not confirm here. The number of undetected sources
in the M83 field excludes those sources from the overlap of the FAUST scan and
the pointed observation.}

\end{deluxetable}

\begin{deluxetable}{ccccc}
\tablecaption{Summary of results}
\small
\tablehead{\colhead{Field} & \colhead{Detections} &
\colhead{Extragalactic} & \colhead{Stellar} & 
\colhead{Unclassified} }
\startdata
Dor & 142 & 13 & 102 & 27  \nl
Cen & 249 & 11 & 159 & 79 \nl 
M83 & 197 & 20 & 169 & 9 \nl
Tel & 189 & 8 & 127 & 54 \nl \hline
Total & 777 & 78 & 509 & 287  \nl

\enddata

\tablecomments{Some of the objects labeled {\bf u} in Table 4 are most likely stars, but
lacking a positive identification as such we keep these in the ``unknown'' bin.}

\end{deluxetable}%

   \newpage

\begin{figure}[tbh]
\vspace{14cm}
\includegraphics{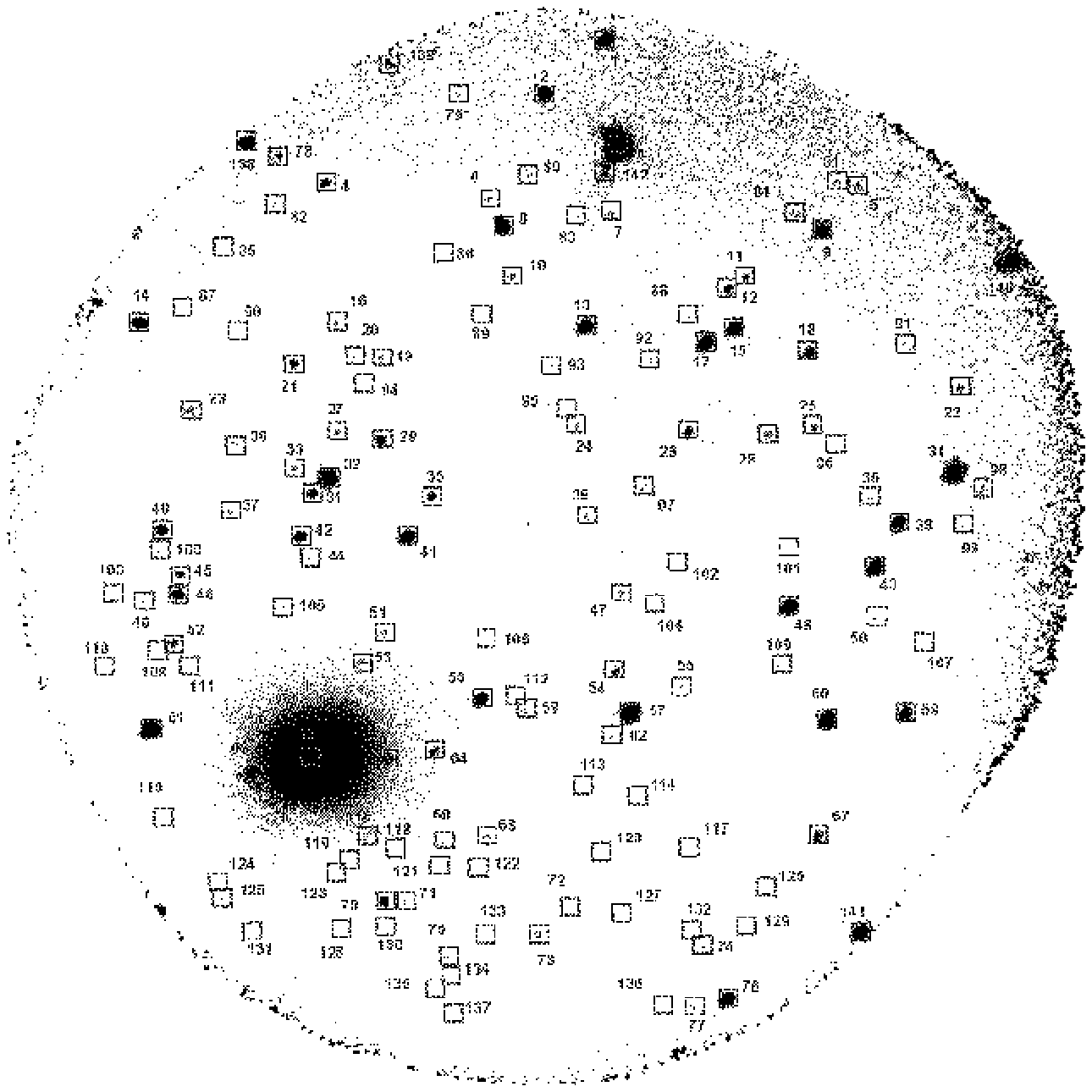}
\end{figure} 
   
\newpage

\begin{figure}[tbh]
\vspace{14cm}
\includegraphics{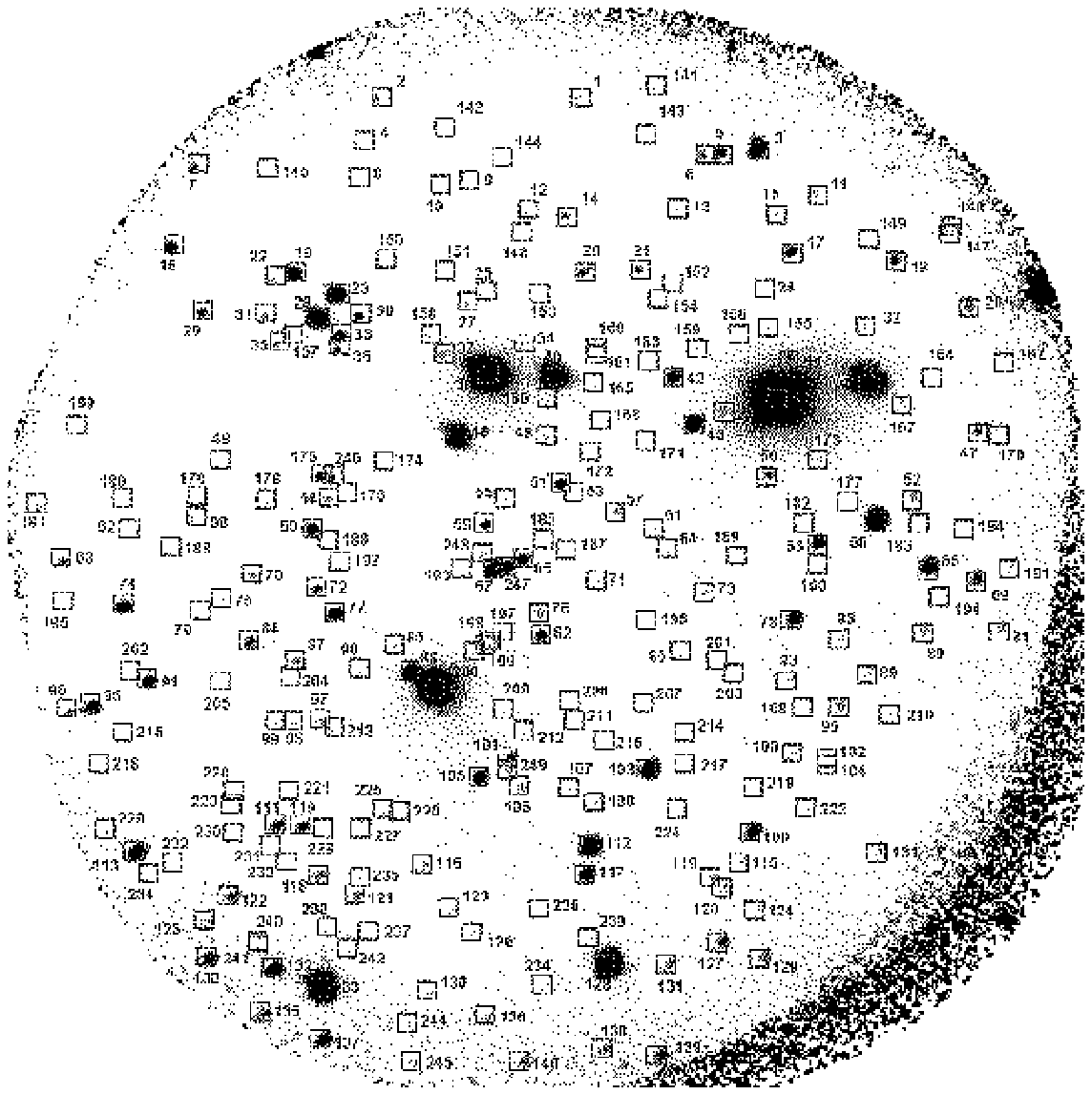}
\end{figure} 
   
\newpage

\begin{figure}[tbh]
\vspace{14cm}
\includegraphics{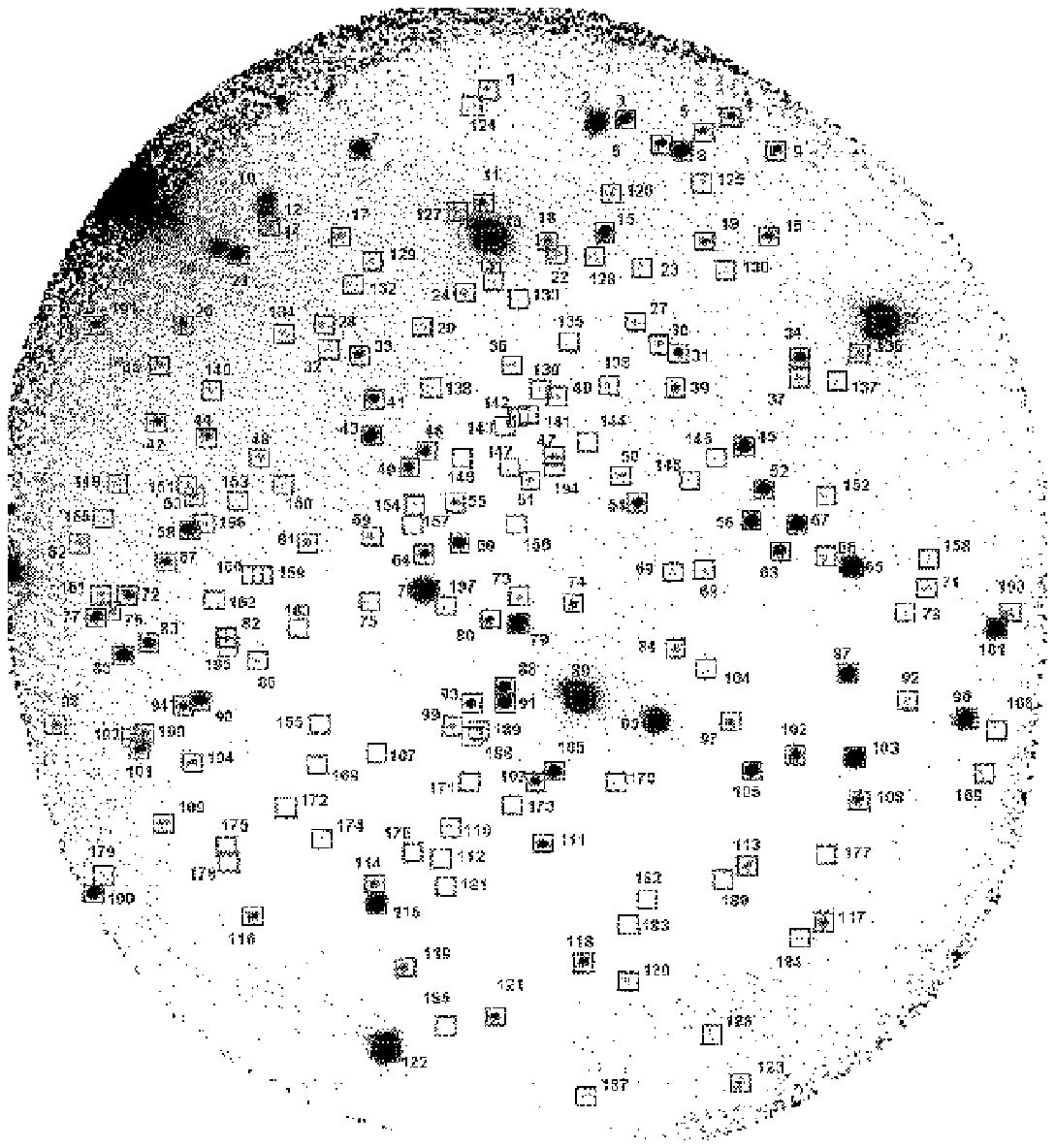}
\end{figure} 
   
\newpage

\begin{figure}[tbh]
\vspace{14cm}
\includegraphics{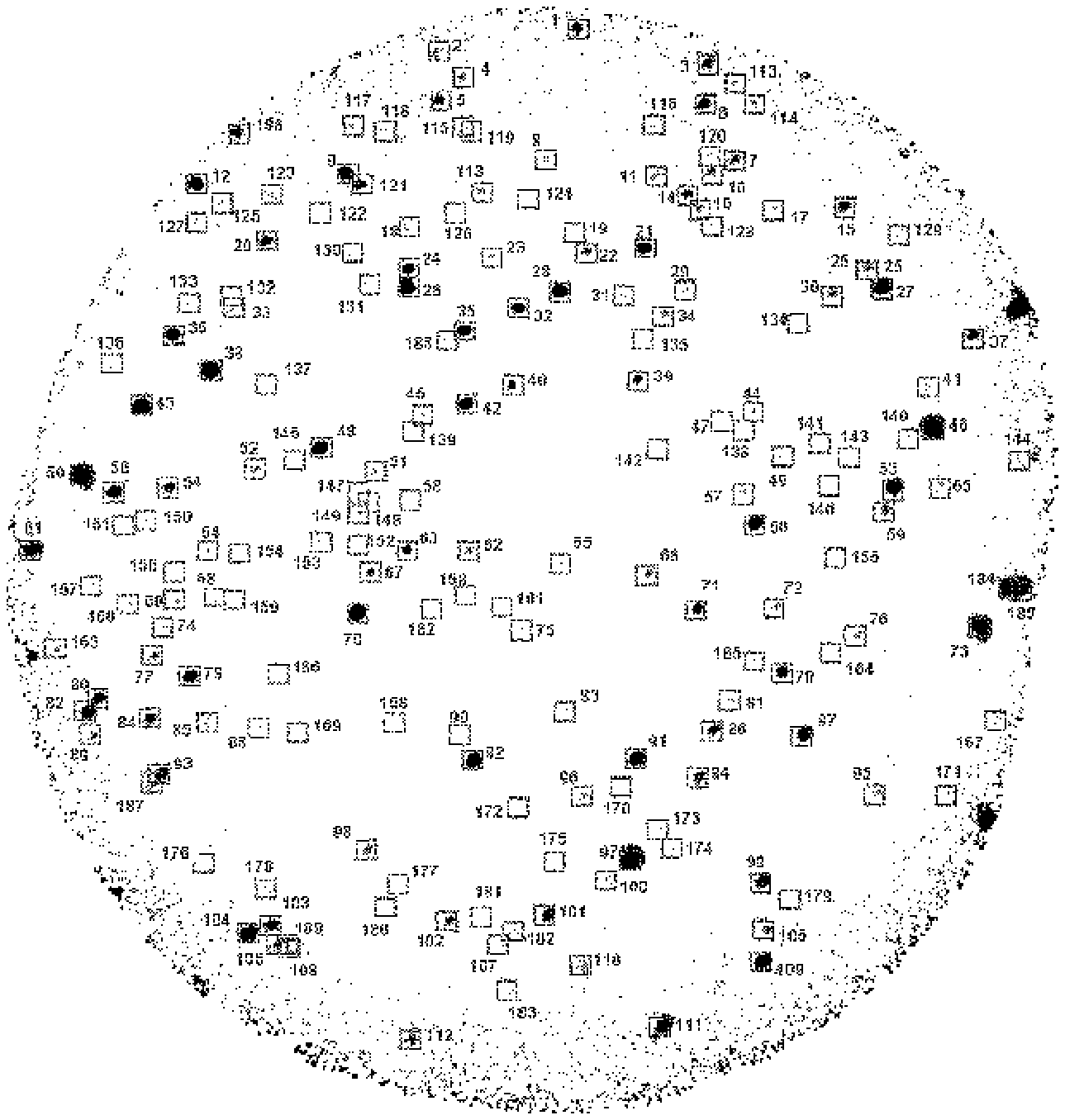}
\end{figure}

\newpage

\begin{figure}
\vspace{15cm}
\putplot{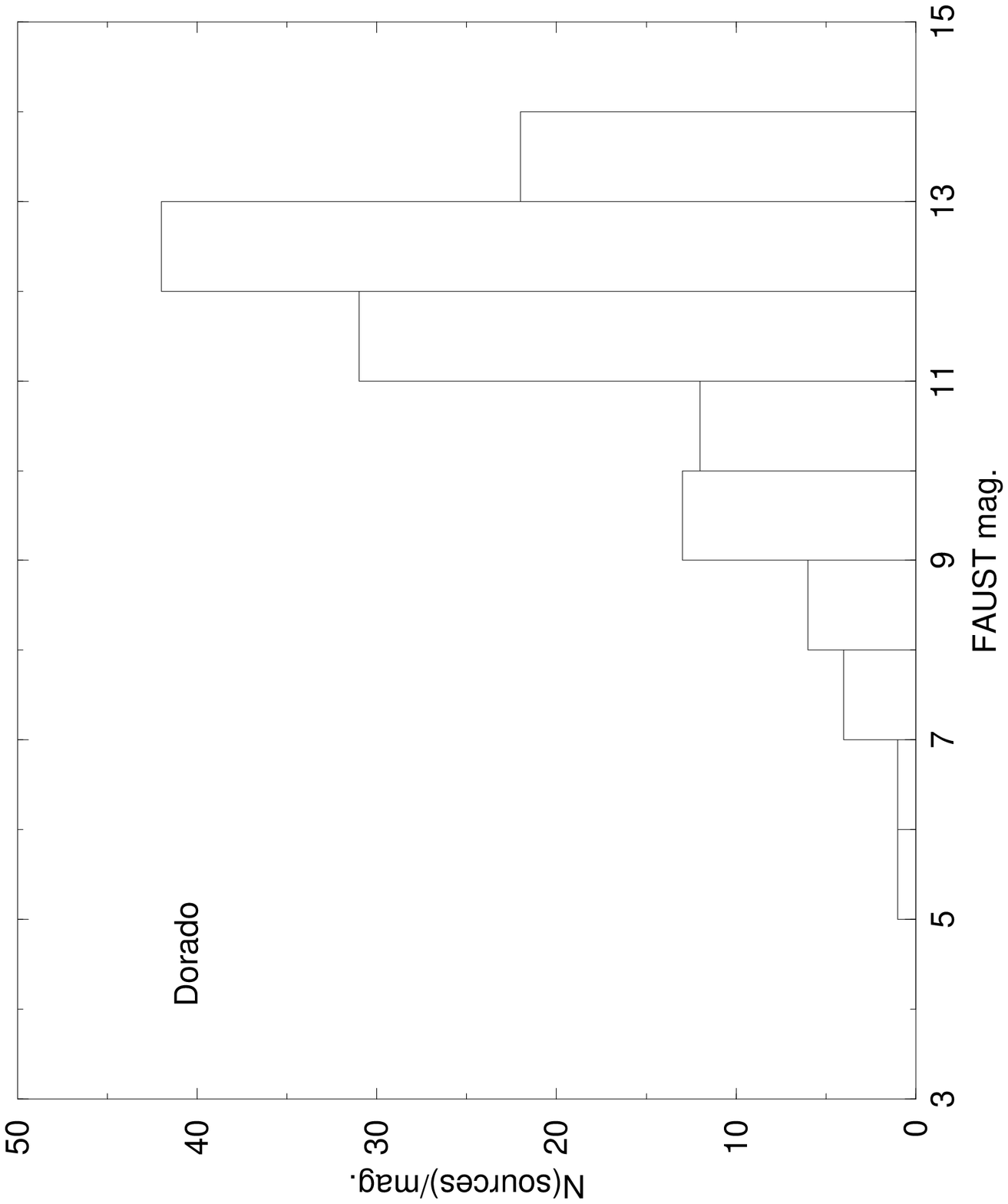}{1in}{-90}{50}{50}{-300}{500}
\putplot{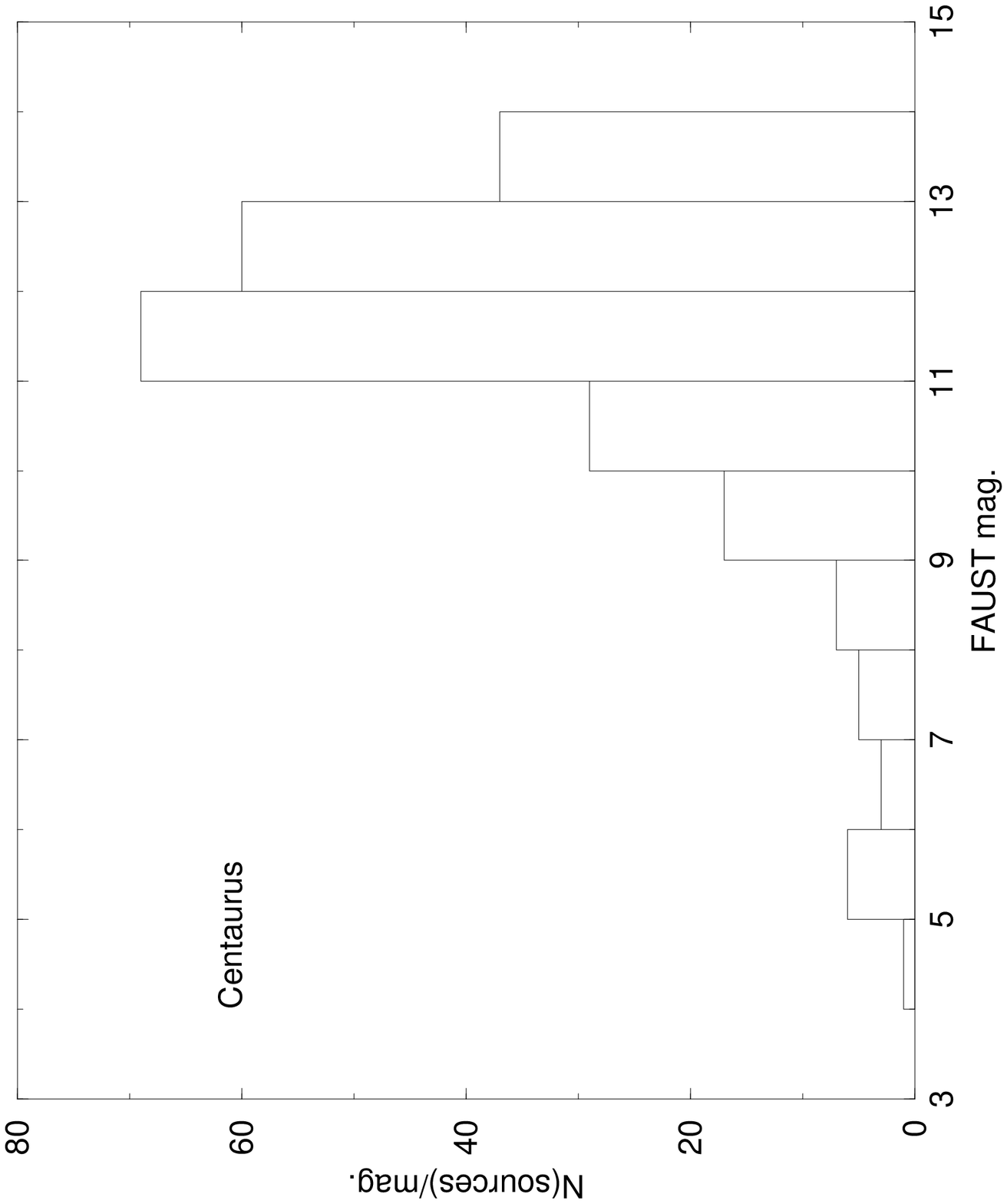}{1in}{-90}{50}{50}{-30}{590}
 \putplot{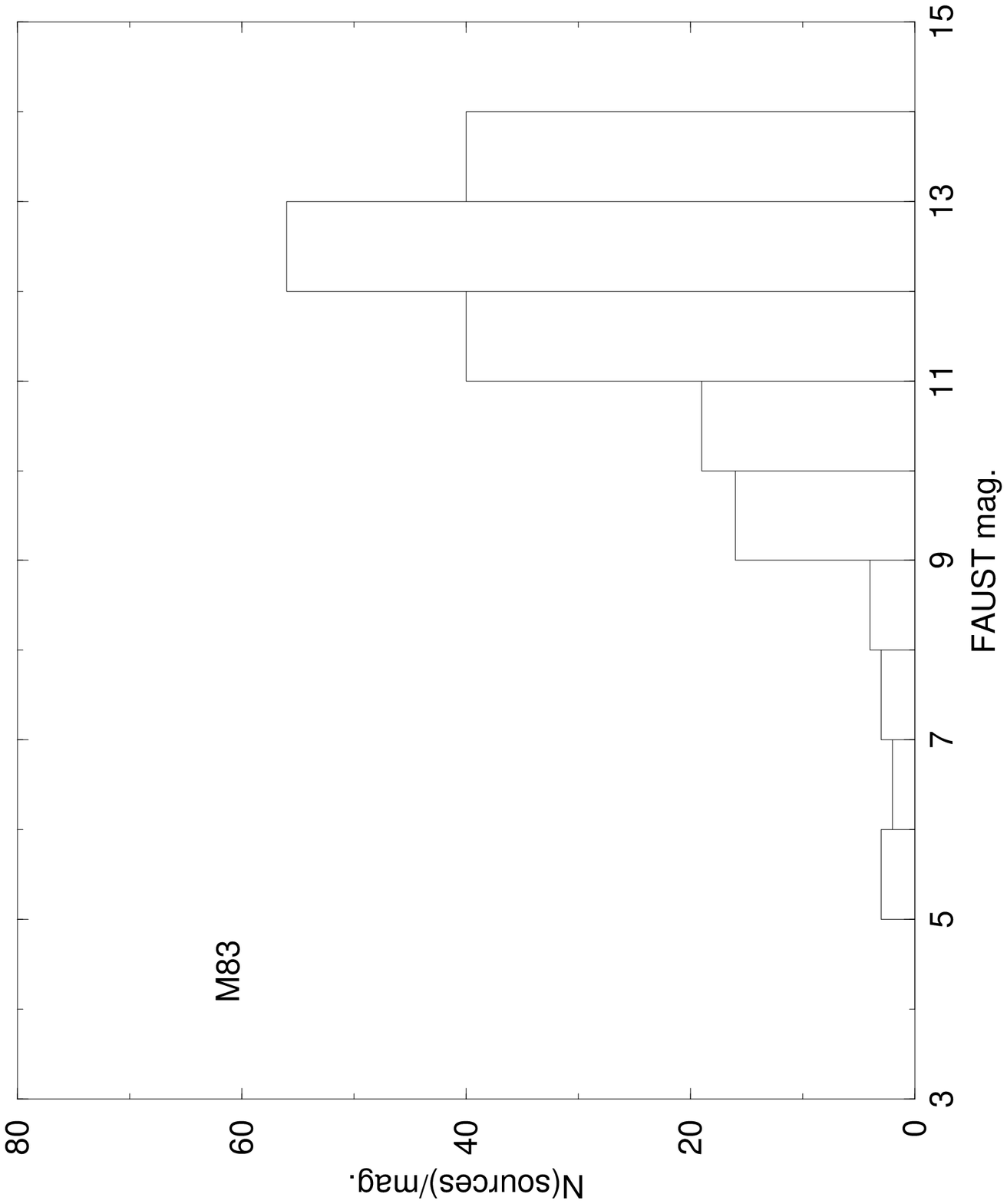}{1in}{-90}{50}{50}{-300}{350}
\putplot{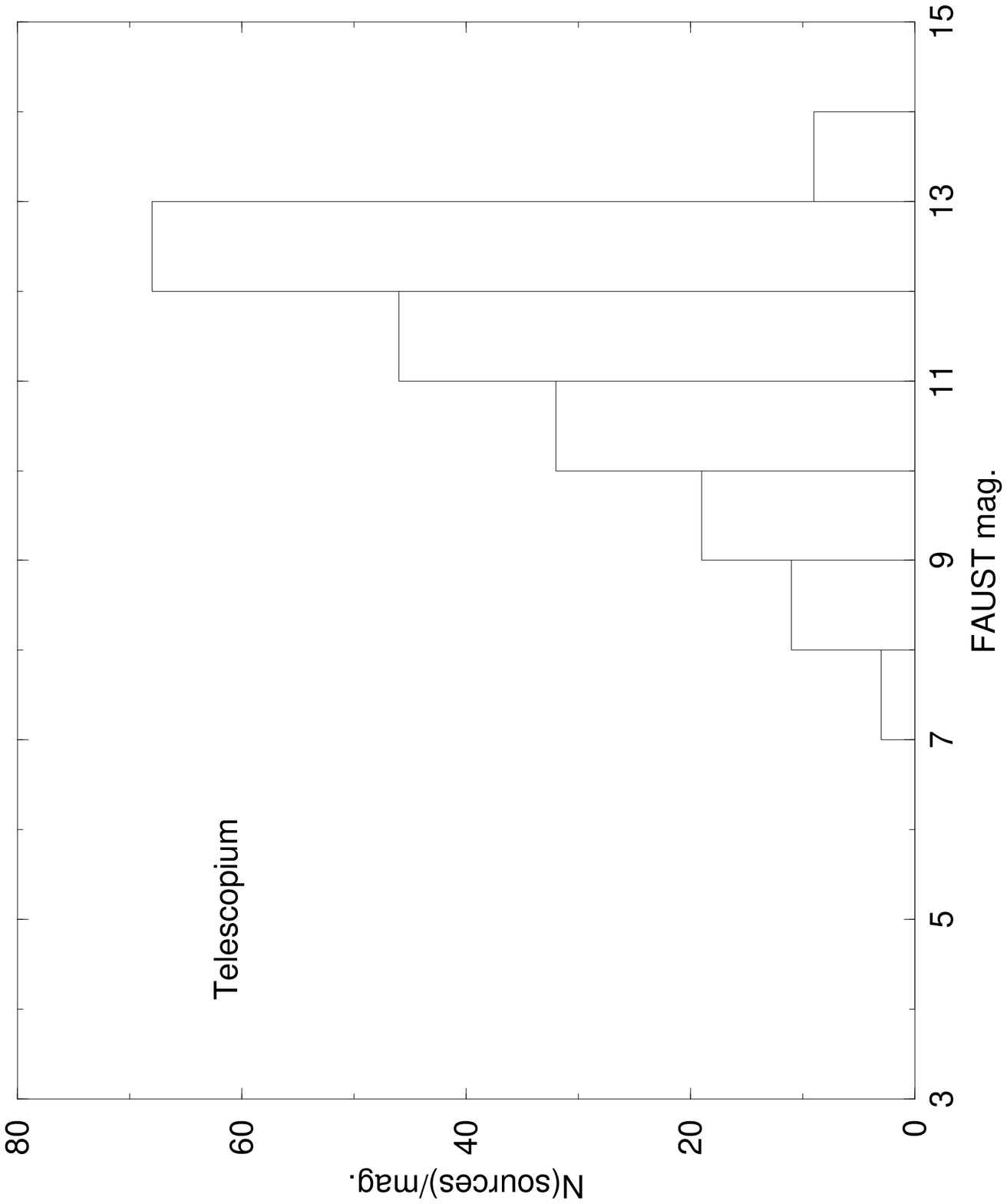}{1in}{-90}{50}{50}{-30}{440}
\end{figure}
\pagebreak
 \newpage

\begin{figure}[tbh]
\vspace{14cm}
\includegraphics{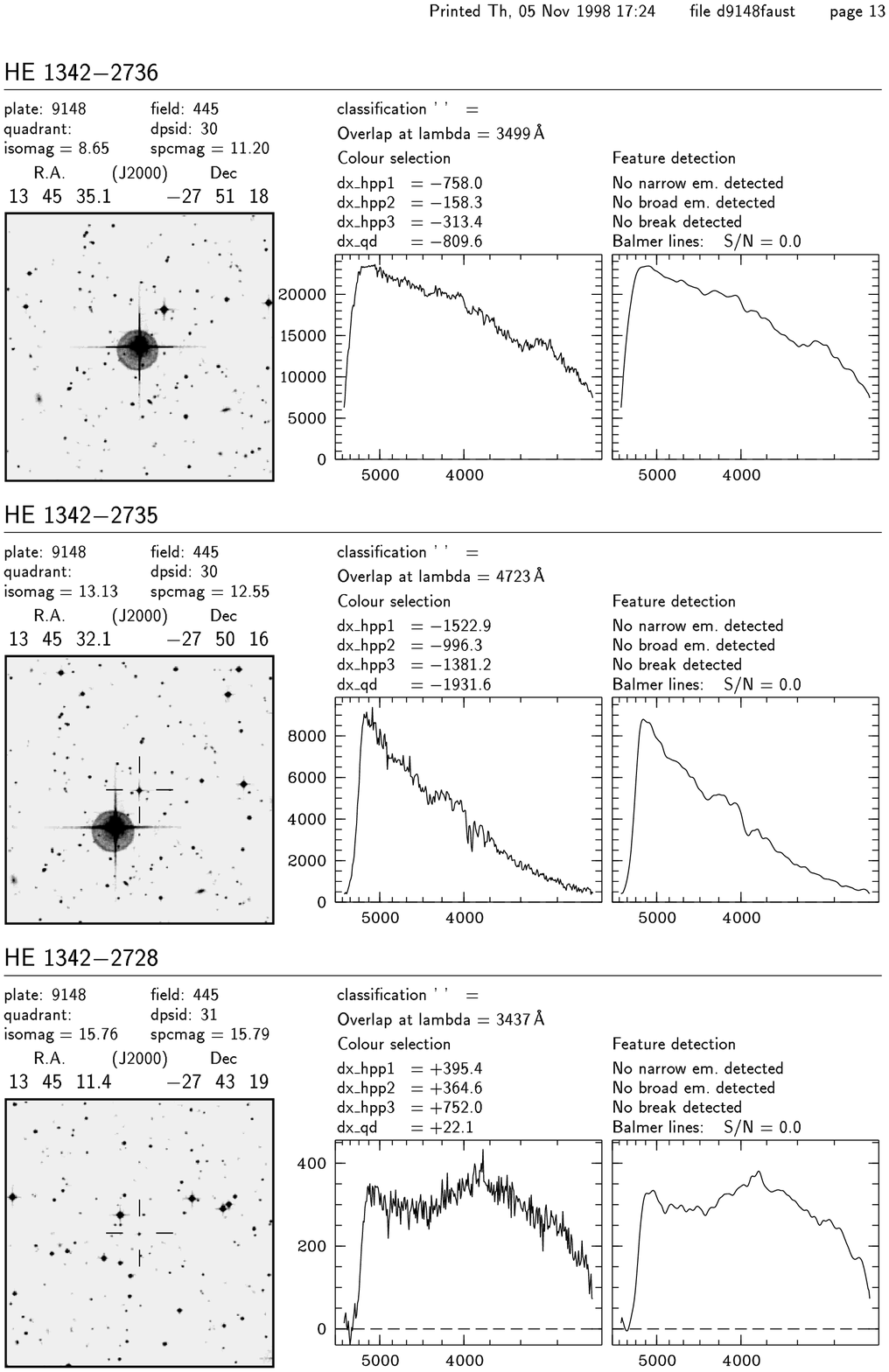}
\end{figure}

\newpage

\begin{figure}
\vspace{15cm}
\putplot{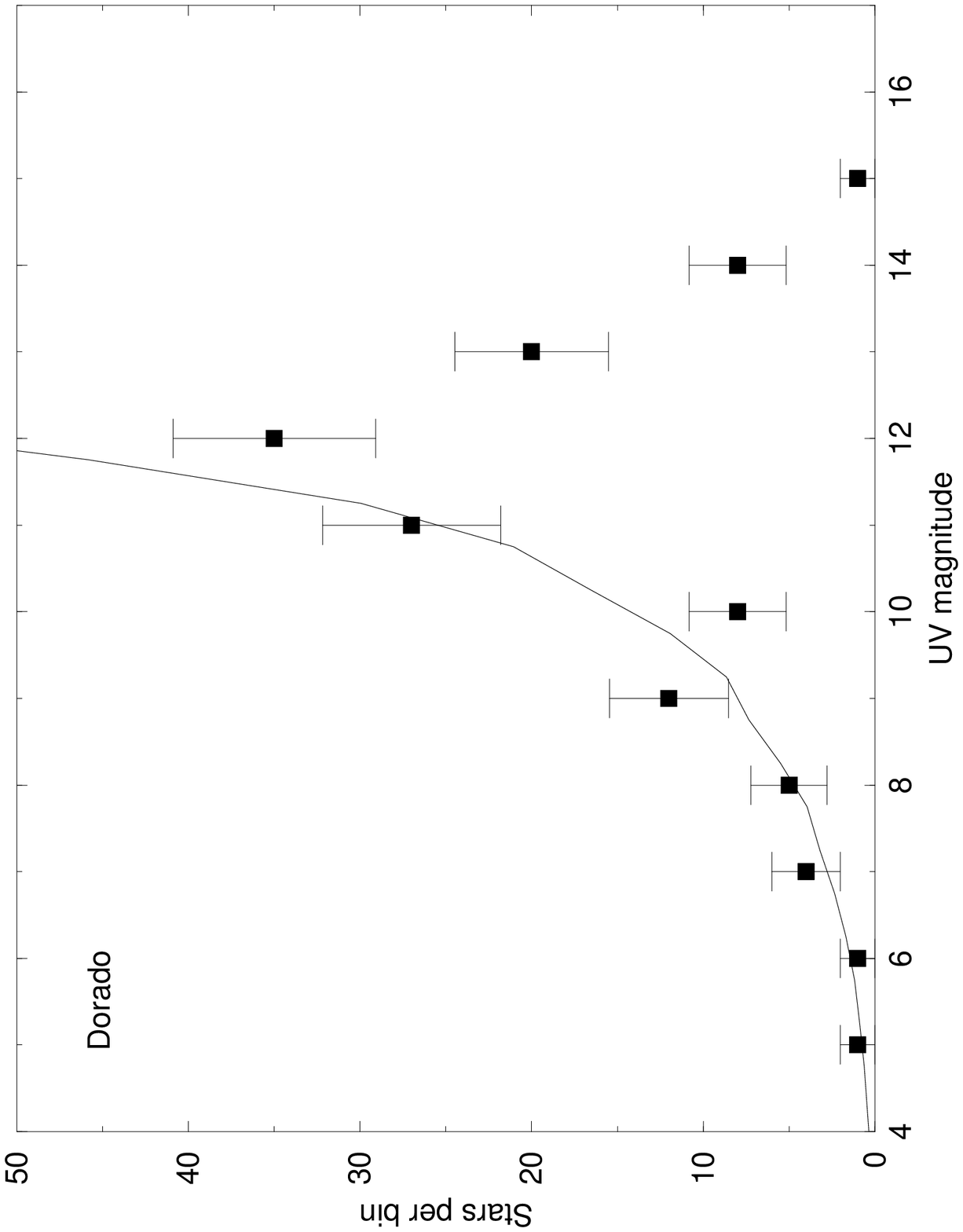}{1in}{-90}{50}{50}{-300}{500}
\putplot{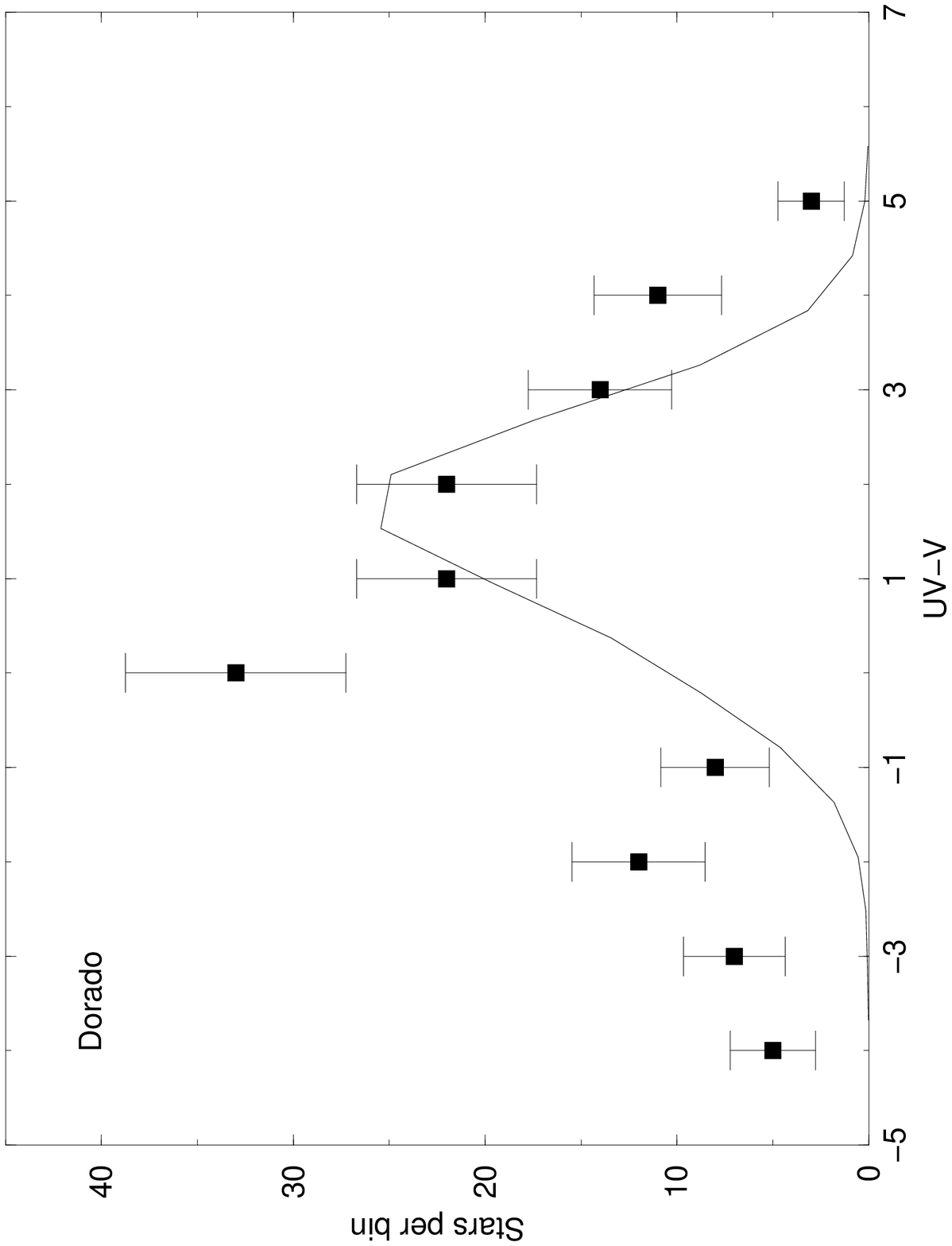}{1in}{-90}{50}{50}{-30}{590}
\putplot{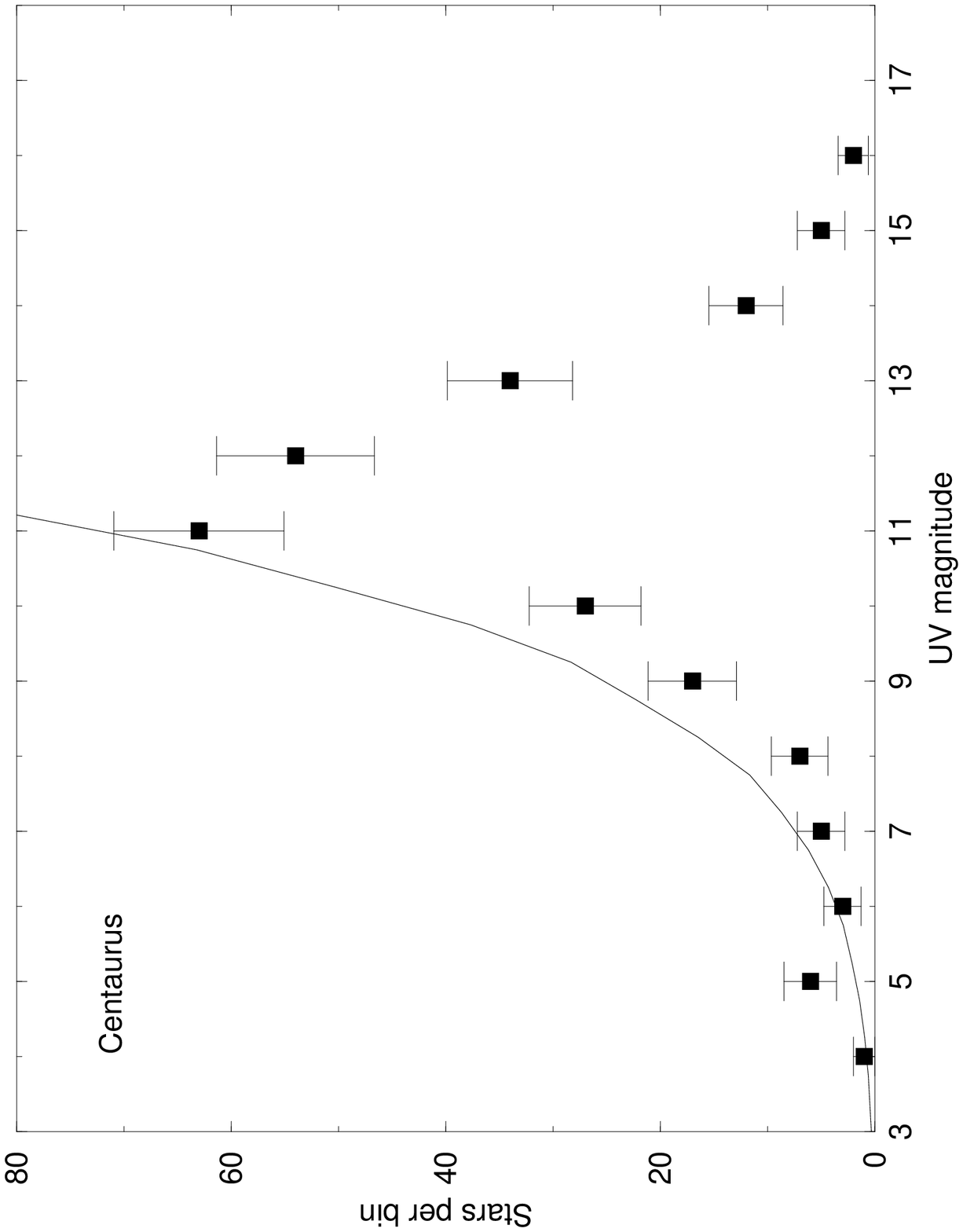}{1in}{-90}{45}{45}{-310}{350}
\putplot{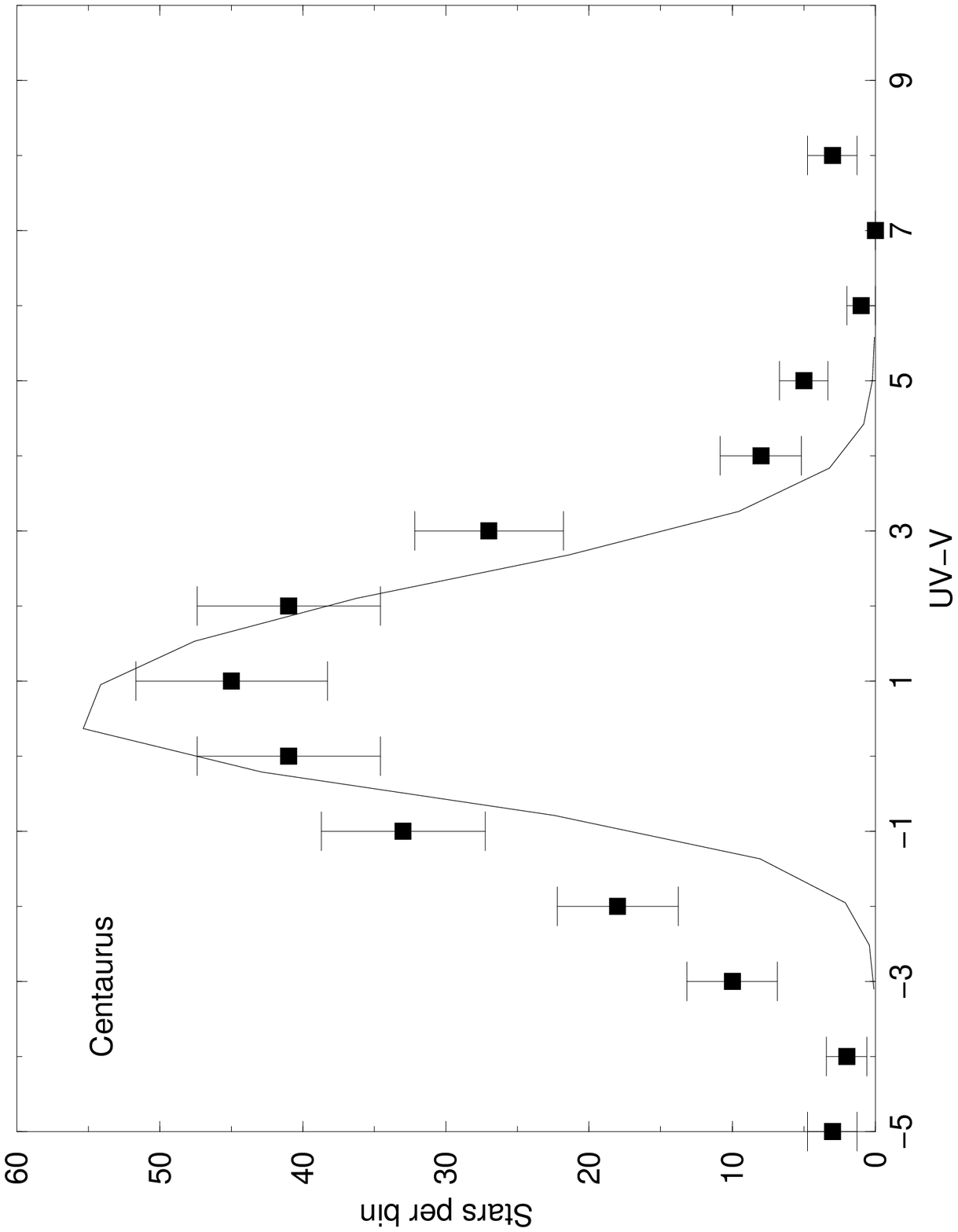}{1in}{-90}{45}{45}{-40}{440}
\end{figure}
\pagebreak
 \newpage
 
\begin{figure}
\vspace{15cm}
\putplot{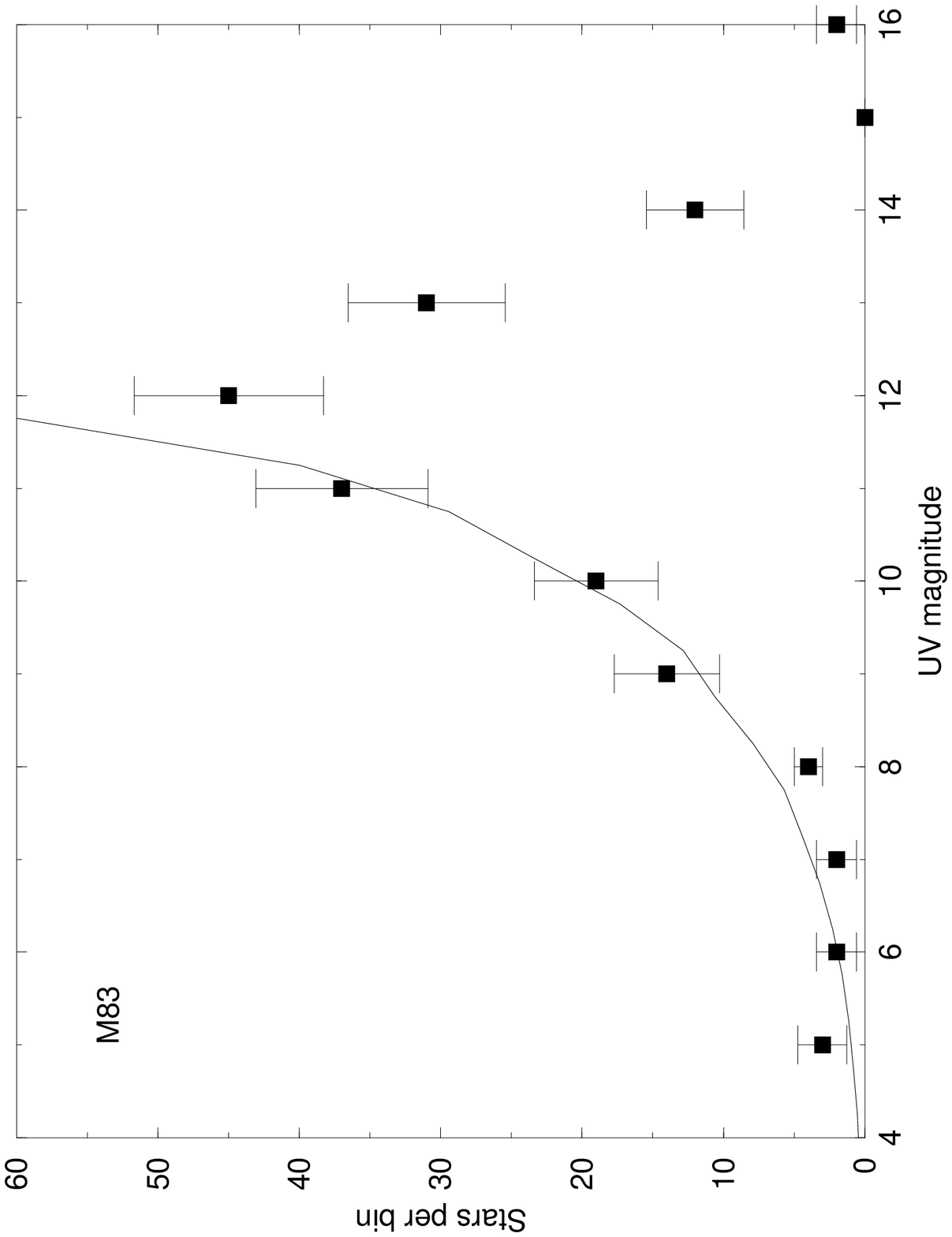}{1in}{-90}{45}{45}{-320}{500}
\putplot{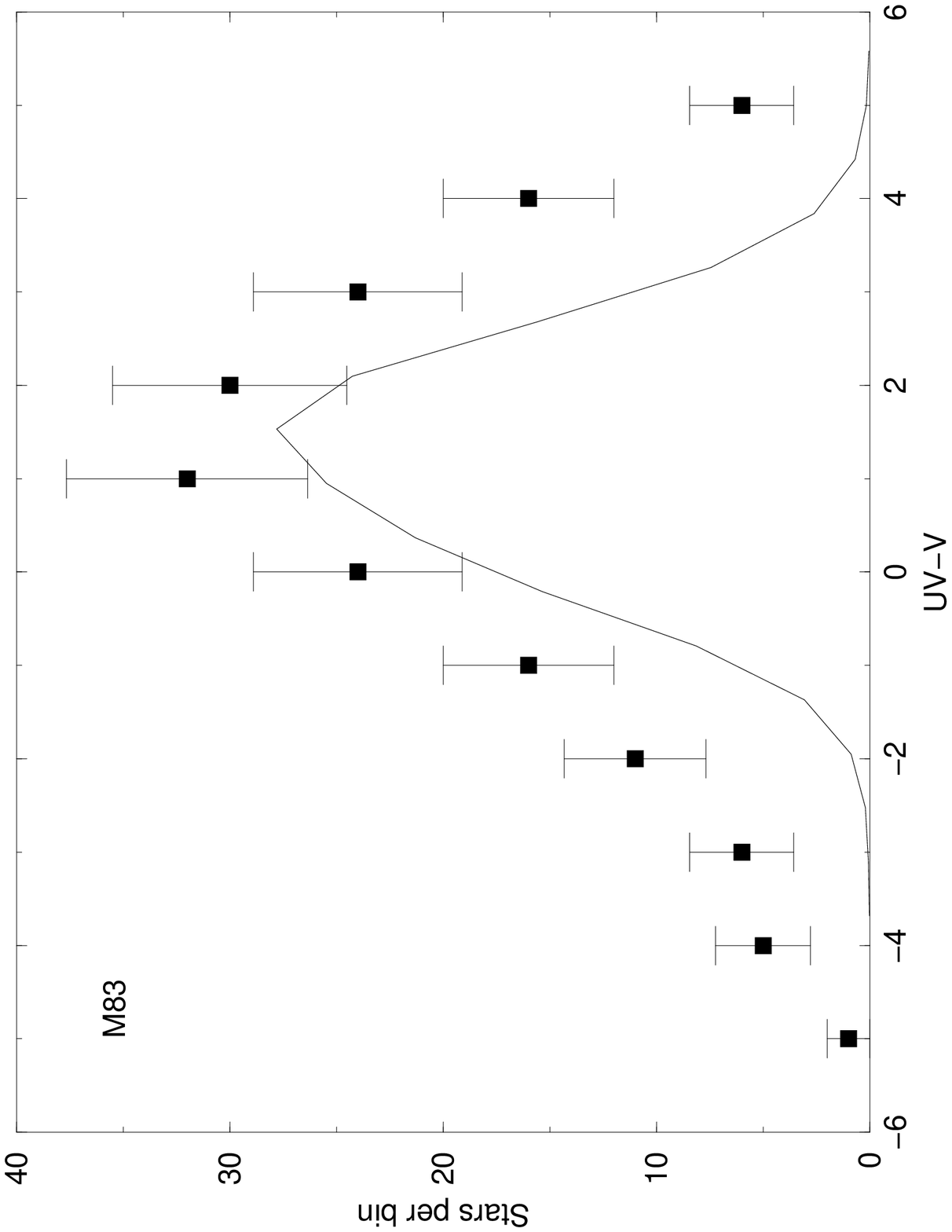}{1in}{-90}{45}{45}{-40}{600}
\putplot{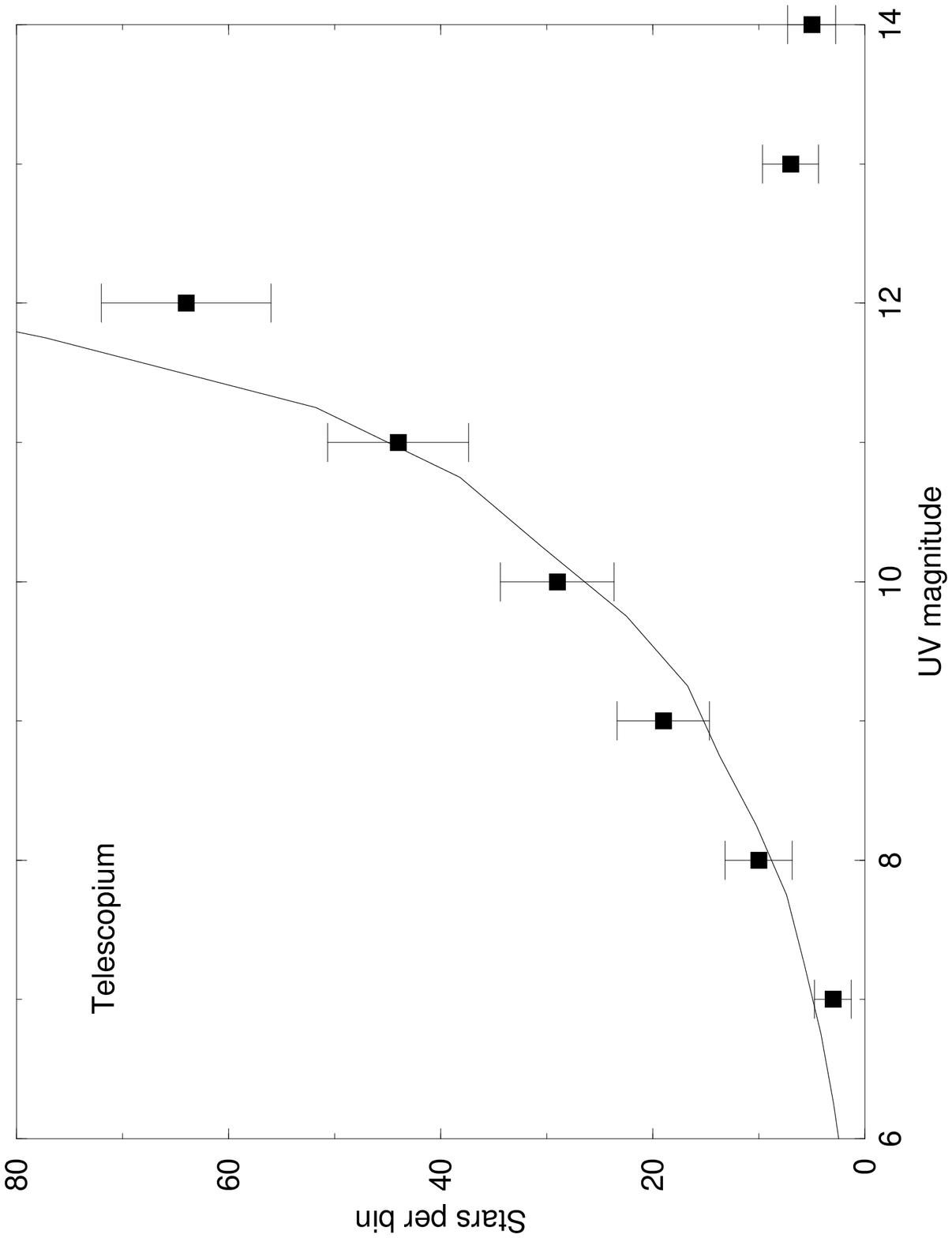}{1in}{-90}{45}{45}{-310}{350}
\putplot{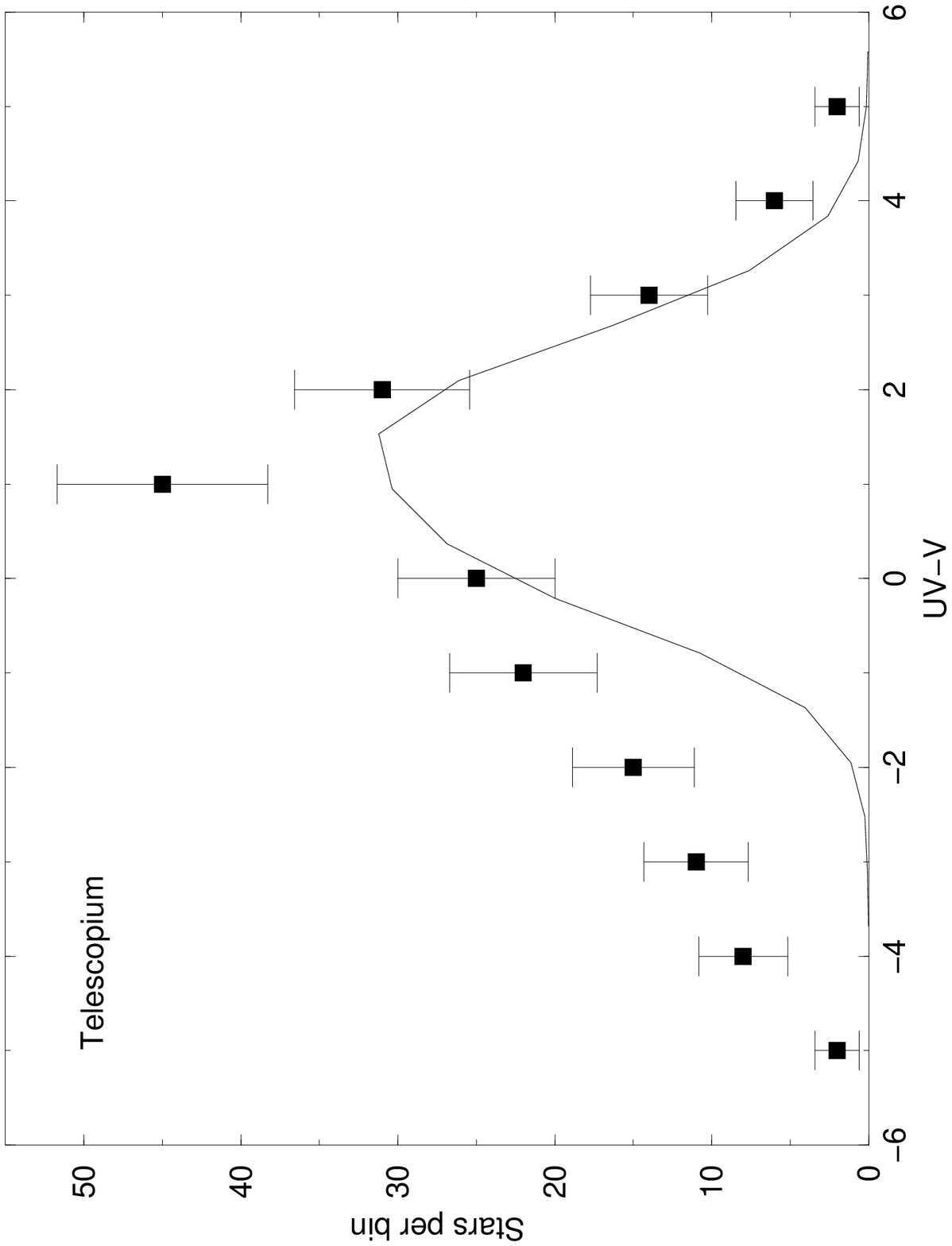}{1in}{-90}{45}{45}{-40}{440}
\end{figure}
\pagebreak
 \newpage

\begin{figure}[tbh]
\vspace{14cm}
\includegraphics{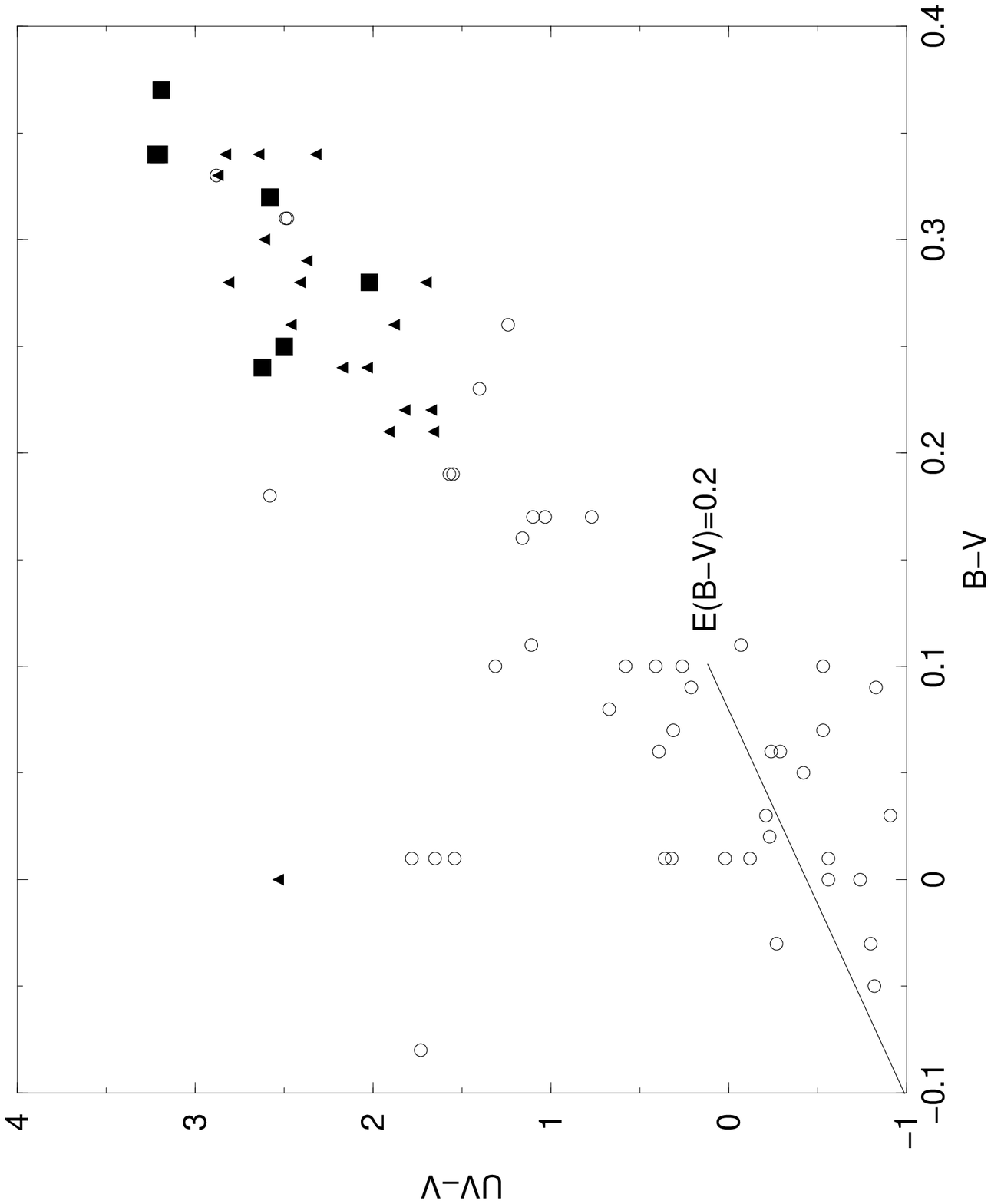}
\end{figure} 
   
\newpage

\end{document}